\documentclass[11pt]{article}
\usepackage{amsmath}  
\usepackage{amsfonts,dsfont}
\usepackage{amssymb}
\usepackage{amsthm}
\usepackage{mathtools}
\usepackage{thmtools}
\usepackage{bm} 
\usepackage{verbatim}
\usepackage{xcolor}
\usepackage{physics}
\usepackage{authblk}
\usepackage{graphicx}
\usepackage[hidelinks]{hyperref}
\usepackage[english]{babel}
\usepackage{float}
\usepackage{tikz}

\newlength{\bredde}
\def\slash#1{\settowidth{\bredde}{$#1$}\ifmmode\,\raisebox{.15ex}{/}
\hspace*{-\bredde} #1\else$\,\raisebox{.15ex}{/}\hspace*{-\bredde} #1$\fi}
\def\cont{
\tikz[baseline=.1ex]{
\draw (0.4ex,-0.99ex+0.6ex) -- (-0.3ex,0+0.6ex);
\draw[->] (-0.3ex,0+0.6ex) -- (0.7ex,0.99ex+0.6ex);
}
}

\newcommand{\RE}{{\rm Re\,}}
\newcommand{\HE}{{\rm He\,}}
\newcommand{\He}{{\rm He\,}}

\newcommand{\IM}{{\rm Im\,}}

\newcommand{\eins}{\leavevmode\hbox{\small1\kern-3.8pt\normalsize1}}
\newcommand{\be}{\begin{equation}}
\newcommand{\ee}{\end{equation}}
\newcommand{\bee}{\begin{eqnarray}}
\newcommand{\eee}{\end{eqnarray}}

\newcommand{\eGinUE}{\textup{eGinUE}}
\newcommand{\eGinOE}{\textup{eGinOE}}
\newcommand{\Ai}{\text{Ai}}

\newtheorem{thm}{Theorem}[section]

\newtheorem{cor}[thm]{Corollary}

\newtheorem{prop}[thm]{Proposition}
\newtheorem*{prob*}{Problem}
\newtheorem*{thm*}{Theorem}

\theoremstyle{definition}

\newtheorem*{defn*}{Definition}
\newtheorem{rem}[thm]{Remark}

\newtheorem*{rem*}{Remark}
\numberwithin{equation}{section}
\newcommand{\erfc}{\text{erfc}}

\providecommand{\keywords}[1]
{
  \small	
  \textbf{Keywords:} #1
}

\title{Spectral density of complex eigenvalues and associated mean eigenvector self-overlaps at the edge of elliptic Ginibre ensembles}
\author{Mark J. Crumpton \& Tim R. W\"urfel}
\affil{Department of Mathematics, King's College London, London WC2R 2LS, UK}

\date{\today}

\begin{document}

\maketitle
\begin{abstract}

\noindent
We consider the density of complex eigenvalues, $\rho(z)$, and the associated mean eigenvector self-overlaps, $\mathcal{O}(z)$, at the spectral edge of $N \times N$ real and complex elliptic Ginibre matrices, as $N \to \infty$. Two different regimes of ellipticity are studied: strong non-Hermiticity, keeping the ellipticity parameter $\tau$ fixed and weak non-Hermiticity with $\tau \rightarrow 1 $ as $N \rightarrow \infty$. At strong non-Hermiticity, we observe that both $\rho(z)$ and $\mathcal{O}(z)$ have the same leading order behaviour across the elliptic Ginibre ensembles, establishing the expected universality. In the limit of weak non-Hermiticity, we find different results for $\rho(z)$ and $\mathcal{O}(z)$ across the two ensembles. This paper is the final of three papers that we have presented addressing the mean self-overlap of eigenvectors in these ensembles.

\end{abstract}



\section{Introduction}\label{sec:intro}

Statistical properties of eigenvalues and eigenvectors have attracted a wealth of attention in the field of Random Matrix Theory (RMT) over recent decades, see reviews such as \cite{Mehta,KS,ForresterBook,BF1,BF2} for a more detailed introduction and more information. Such spectral studies are well established in the case of Hermitian matrices, which are equal to their own adjoint. Therein, eigenvalues are purely real and the eigenvectors form an orthogonal basis, leading to trivial correlations between eigenvectors, thus focusing attention to eigenvalues only. Celebrated results for Hermitian random matrices include, for example, the Tracy-Widom distribution \cite{TW} and the Wigner semi-circle law \cite{Wigner}. However, such results are much harder to obtain in the realm of non-Hermitian matrices, due to the presence of non-trivial correlations between eigenvectors, which are not constrained further by rotational symmetry. 

Initially, we focus on the eigenvalues, $z_n$, of $N \times N$ non-Hermitian matrices, which are not solely constrained to the real line, but may take complex values. The most natural object to study is $\rho (z)$, the mean spectral density at a point $z$ in the complex plane, which is defined as
\begin{equation}
    \rho\, (z) \equiv \bigg\langle \frac{1}{N}\sum_{n=1}^N \delta(z-z_n) \bigg\rangle \ , 
    \label{eq:rho_def}
\end{equation}
where $\delta(\cdot)$ is the Dirac $\delta-$function and $\langle \cdot \rangle$ stands for the expectation value with respect to the joint probability density function (jpdf) of a particular random matrix ensemble. Popular ensembles within which one can study this density are the \emph{Ginibre ensembles} \cite{Ginibre}, featuring independent, identically distributed (iid) Gaussian entries. These entries can be real, complex or even quaternion and the corresponding ensembles are abbreviated as GinOE, GinUE and GinSE respectively. For such matrices it is well-known that, in the limit of large-$N$, the complex eigenvalues converge to a uniform disc in the complex plane \cite{Ginibre}, with less trivial density behaviour at the edge of this disc \cite{Rider,BS}. One can extend these ensembles into \emph{elliptic Ginibre ensembles} - abbreviated as eGinOE, eGinUE and eGinSE respectively - through the introduction of correlations between off-diagonal elements, using a single parameter $\tau\in[0,1)$ \cite{Girko1, Girko2}. In such ensembles, as $N$ becomes large, the complex eigenvalues are now constrained to an elliptic droplet, with dimensions dependent upon $\tau$. 

Focusing on complex eigenvalues of eGinOE and eGinUE matrices, finite-$N$ and asymptotic results for the mean spectral density and other statistical properties of eigenvalues are well-known for much of the complex plane, see e.g. \cite{KS,BF1,BF2,FN,Bender,AB,AP,LR,GGNV,CFW24}. Typically, large-$N$ asymptotic results are calculated for two regimes of $\tau$, these are the limits of \emph{strong non-Hermiticity}, with fixed $\tau$, and \emph{weak non-Hermiticity}, where $\tau \to 1$ as $N \to \infty$, as introduced in \cite{FKS97a,FKS97b,FKS98}. The density of complex eigenvalues at the edge of the elliptic droplet has previously been considered at strong and weak non-Hermiticity for the eGinOE \cite{AP}. Additionally, the corresponding results exist in the eGinUE due to the work of Lee and Riser \cite{LR} for strong non-Hermiticity and Garci\'{a}-Garci\'{a} et. al. \cite{GGNV} for weak non-Hermiticity, as well as \cite{Bender, AB}. In the limit of strong non-Hermiticity, universality is expected among all members of the class of real and complex non-Hermitian matrices with iid entries, in the bulk and at the edge of the droplet \cite{TaoVu}. See also \cite{Osman23, OM23, Osman24} for more recent works on universality of eigenvalue correlation functions in the bulk. In the present paper, among other tasks, we aim to provide additional insights and numerical evidence of this feature at the spectral edge of real and complex elliptic Ginibre matrices.

For real Ginibre-type matrices, a macroscopic proportion of the eigenvalues are constrained to the real line and so one must consider the density of complex and real eigenvalues separately. The latter having already been studied at finite-$N$ and in several asymptotic regimes in the GinOE and eGinOE, see \cite{FN,FN08,Tarnowski24} for more. Additionally, the spectral density of the induced extensions of the real and complex Ginibre ensembles has been studied at finite-$N$ and in various large-$N$ limits by Fischmann et al. \cite{FBKSZ}. There, for both ensembles, they observed a bulk spectrum of eigenvalues existing in an annulus as $N \to \infty$, similar to what was observed in \cite{FeinbergZee97, GKZ2011}. Annular spectra are also a feature of the ensemble of sub-unitary matrices studied at finite-$N$ and asymptotically by Wei and Fyodorov and Bogomolny respectively \cite{WF, Bogomolny}. 

On the other hand, the study of statistical properties of eigenvectors is also an interesting avenue. In general, the eigenvalues of a random matrix $X$ have multiplicity one, thus one can safely write that $X = S \Lambda S^{-1}$. Here, $\Lambda$ is a diagonal matrix containing the eigenvalues of $X$ and the columns (rows) of $S$ $(S^{-1})$ are the right (left) eigenvectors of $X$, denoted as $\bm v_{R,i}$ and $\bm v_{L,i}^\dagger$, which satisfy $\bm v_{L,i}^\dagger X = z_i \bm v_{L,i}^\dagger$ and $X \bm v_{R,i} = z_i \bm v_{R,i}$ respectively. Within this setup, the left and right eigenvectors obey a bi-orthogonality condition, such that $\bm v_{L,i}^\dagger \bm v_{R,i} = \delta_{ij}$. However, providing that $X$ is not normal (i.e. it does not commute with its own adjoint), a relationship of this nature does not exist within the sets of left and right eigenvectors. As such, it is the relations within sets of left and right eigenvectors that can be used as a measure of non-normality. In order to quantitatively study this effect, Chalker \& Mehlig introduced the now well-known matrix of overlaps with entries
\begin{equation}
    \mathcal{O}_{mn} = \left( \bm v^\dagger_{L,m} \bm v_{L,n} \right) \left( \bm v^\dagger_{R,n} \bm v_{R,m} \right) \quad m, n=1, 2, \ldots, N \ ,
    \label{eq:def_overlap}
\end{equation}
such that the so-called \textit{self-overlaps}, $\mathcal{O}_{nn}$, are the diagonal entries of this matrix \cite{CM,MC}. Within their works, they studied the large-$N$ asymptotic limits of the connected ensemble averages
\begin{equation}
    \mathcal{O}(z) \equiv \bigg\langle \frac{1}{N}\sum_{n=1}^N \mathcal{O}_{nn} \ \delta(z-z_n) \bigg\rangle \hspace{0.75cm}  \text{and} \hspace{0.75cm} \mathcal{O}(z_1,z_2) \equiv \bigg\langle \frac{1}{N^2}\sum_{\substack{m,n=1 \\ m \neq n}}^N \mathcal{O}_{mn} \ \delta(z_1 -z_m) \  \delta(z_2-z_n) \bigg\rangle\ ,
    \label{eq:O_def}
\end{equation}
finding that, within the spectral bulk of the GinUE, $\mathcal{O}(z) \sim N (1 - |z|^2)/\pi$ and zero otherwise. This showed that the mean self-overlap in non-normal matrices is macroscopically larger than the corresponding value in normal matrices, where $\mathcal{O}_{nn} = 1$ for all $n=1,\ldots,N$. 

The findings of Chalker \& Mehlig prompted further study of the self-overlap in a few other ensembles of random non-normal matrices. For example, in subsequent years they extended their own work to consider the eGinUE \cite{MC} and Janik et al. introduced a new method to calculate the mean self-overlap from the spectral Green's function \cite{JNNPZ}, independently verifying the results of Chalker \& Mehlig. Recently, this field received a further boost in attention as the papers by Bourgade \& Dubach \cite{BD} and Fyodorov \cite{FyodorovCMP} simultaneously extended the work of Chalker \& Mehlig to obtain the full distribution of the mean self-overlap, for both finite-$N$ and in various asymptotic limits. Additionally, Bourgade \& Dubach used techniques from free probability to derive statistics of the off-diagonal overlap, whereas Fyodorov utilised incomplete Schur decomposition \cite{Edelman} to derive the distribution of the self-overlap associated with real eigenvalues in the GinOE. These papers led to a surge in interest in the self-overlap of eigenvectors in Ginibre ensembles with results that include: the distribution of the self-overlap associated with real eigenvalues in the eGinOE at finite-$N$ and in different asymptotic regimes \cite{Tarnowski24,FT}, the mean self-overlap associated with complex eigenvalues in the GinOE, eGinUE and eGinOE \cite{CFW24,WCF23} and the mean self-overlap and off-diagonal overlap in the GinSE \cite{AFK}. Other results in the study of eigenvectors of random matrices and the self-overlap include \cite{BZ, ATTZ, WS, FSav2, FyoOsm22, Noda23a, Noda23b}. For a broader review of results and applications of the overlap matrix, one can consult \cite{CFW24,WCF23} and references therein, which lay the groundwork for the present paper.

In spite of recent developments, to the best of our knowledge, results have yet to be derived for the mean self-overlap associated with complex eigenvalues at the edge of both the eGinUE and eGinOE at strong and weak non-Hermiticity. Therefore, the main aim of this paper is to provide results in these limits for both ensembles. Note that edge statistics are an important area of study as they smoothly interpolate the statistics of eigenvalues inside and outside of the bulk, whilst also having applications to chaotic scattering \cite{HILSS, SFPB, FSPB}. As part of our presentation of such results, it is convenient to introduce the \textit{mean conditional self-overlap}
\begin{equation}
    \mathbb{E}\left( z  \right) \equiv \mathbb{E}\left(\mathcal{O}_{nn} \, \vert \, z = z_n \right) = \frac{\mathcal{O}(z)}{ \rho\, (z) } \ ,
    \label{eq:E_def}
\end{equation}
as this is the object that we will mostly be concerned with when making comparisons to numerical data. This object is particularly useful in this work as it serves to verify results for both the density of complex eigenvalues and the associated mean self-overlap of eigenvectors.

The remainder of this manuscript is arranged as follows. In Section \ref{sec:rems_eGinOUE} we provide the necessary mathematical definitions of the eGinOE and eGinUE, whilst reviewing some relevant existing results and important terminology. Then, in Section \ref{sec:Main_Results}, we outline existing results for the density of complex eigenvalues and present our main results for the mean self-overlap associated with such eigenvalues in the eGinUE and eGinOE. This is done at the edge of the elliptic droplet for strong and weak non-Hermiticity in Section \ref{sec:SNH_edge_results} and Section \ref{sec:WNH_edge_results} respectively. We then proceed to review some open problems in Section \ref{sec:open_probs} before, finally, in Section \ref{sec:proofs_eGinOUE_edge} we provide proofs of our main results.

\section{Statement \& Discussion of Main Results}

\subsection{Remarks on Elliptic Ensembles}
\label{sec:rems_eGinOUE}

Complex-valued matrices within the eGinUE have the following jpdf with respect to the flat Lebesgue measure, $dX = \prod_{i,j=1}^N dX_{ij} d\bar{X}_{ij}$, 
\begin{equation}
    \mathcal{P}_{\text{eGinUE}}(X)\,dX = \frac{1}{D_{N,\tau}} \exp{ -\frac{1}{1-\tau^2} \Tr \left( XX^\dagger -\tau \ \RE X^2 \right)} dX \ ,
    \label{eq:jpdf_eGinUE}
\end{equation}
with normalisation constant $D_{N,\tau} = \pi^{N^2} \left( 1-\tau^2 \right)^{\frac{N^2}{2}}$ \cite{Girko2,SCSS}. On the other hand, real-valued matrices in the eGinOE have the following jpdf with respect to the flat real Lebesgue measure 
\begin{equation}
    \mathcal{P}_{\text{eGinOE}}(X)\,dX = \frac{1}{C_{N,\tau}} \exp{-\frac{1}{2(1-\tau^2)} \Tr \left( XX^T -\tau X^2 \right)} dX \ ,
    \label{eq:jpdf_eGinOE}
\end{equation}
with normalisation constant $C_{N,\tau} = (2\pi)^{N^2/2} \left( 1+\tau \right)^{\frac{N(N+1)}{4}} \left( 1- \tau \right)^{\frac{N(N-1)}{4}}$ \cite{Girko2,FT}. In both the eGinUE and eGinOE the parameter $\tau$ quantifies the correlation between pairs of off-diagonal elements, with $\tau=0$ corresponding to uncorrelated GinUE or GinOE matrices and $\tau = 1$ yielding Hermitian or real symmetric matrices. The parameter $\tau$ also controls the shape of the droplet to which the eigenvalues converge in the complex plane. For large-$N$, the eGinOE and eGinUE have been studied in two distinctly different regimes of $\tau$, namely strong and weak non-Hermiticity, abbreviated as SNH and WNH respectively. At SNH, $\tau \in [0,1)$ and remains fixed as $N \to \infty$, whereas at WNH, $\tau$ asymptotically approaches unity as $N$ becomes large, via $\tau = 1 - (\pi \alpha)^2/2N$, where $\alpha$ remains finite.

The density of complex eigenvalues at finite-$N$ has previously been found in both the eGinUE and eGinOE, see \cite{KS,BF1} for the eGinUE and \cite{FN08,APS} for the eGinOE. In the eGinUE, the result is given by
\begin{equation}
    \rho^{(\text{eGinUE},c)}_N(z) = \frac{1}{\pi} \frac{1}{\sqrt{1-\tau^2}}\exp \left\{ -\frac{\vert z \vert^2 - \tau \ \text{Re}(z^2)}{1-\tau^2} \right\} \sum_{k=0}^{N-1} \frac{\tau^k}{k!}\HE_k\left( \frac{z}{\sqrt{\tau}} \right)\HE_k\left( \frac{\bar{z}}{\sqrt{\tau}} \right) \ ,
    \label{eq:rho_N_eGinUE}
\end{equation}
and correspondingly, in the eGinOE, 
\begin{equation}
    \rho_N^{(\text{eGinOE,c})}(z)= \sqrt{\frac{2}{\pi}}\frac{\Im(z)}{1+\tau}  \ \exp \left\{ \frac{\Im(z)^2 - \Re(z)^2}{1+\tau} \right\} \text{erfc}\left( \sqrt{\frac{2}{1-\tau^2}} \ \vert \Im(z) \vert \right) P_{N-2} \ ,
    \label{eq:rho_N_eGinOE}
\end{equation}
such that
\begin{equation}
     P_N =  \frac{1}{\bar{z}-z} \sum_{k=0}^{N} \ \frac{\tau^{k+\frac{1}{2}}}{k!} \bigg[ \HE_{k+1}\left(\frac{\bar{z}}{\sqrt{\tau}}\right) \HE_{k}\left( \frac{z}{\sqrt{\tau}} \right) - \HE_{k+1}\left( \frac{z}{\sqrt{\tau}} \right) \HE_{k}\left(\frac{\bar{z}}{\sqrt{\tau}}\right) \bigg] \ .
     \label{eq:P_N}
\end{equation} 
Note that in the above equations we have introduced the Hermite polynomials 
\begin{equation} 
    \HE_k(x) = \frac{1}{\sqrt{2\pi}} \int_{-\infty}^{\infty} dy \, \exp{-\frac{1}{2}y^2} \, \left( x+iy \right)^k = k! \sum_{m=0}^{[k/2]} \frac{(-1)^m}{(k - 2m)! m!} \frac{x^{k - 2m}}{2^m} 
    \label{eq:Hermite_Poly}
\end{equation}
and the complementary error function, $\erfc(x) = 1 - \erf(x)$, where $\erf(x) = \frac{2}{\sqrt{\pi}} \int_0^x e^{-t^2} dt$. One should also note that the density of real eigenvalues in the eGinOE is known at finite-$N$ and in various asymptotic regimes, see \cite{FN08,ForM} for results at SNH and \cite{FKS98,Tarnowski24,Efe97} for WNH. In this manuscript, we will not discuss the density of real eigenvalues, or associated self-overlaps further, instead, we will focus on complex eigenvalues and their eigenvector self-overlaps. 

\begin{figure}[h!]
    \centering
    \includegraphics[scale = 0.335]{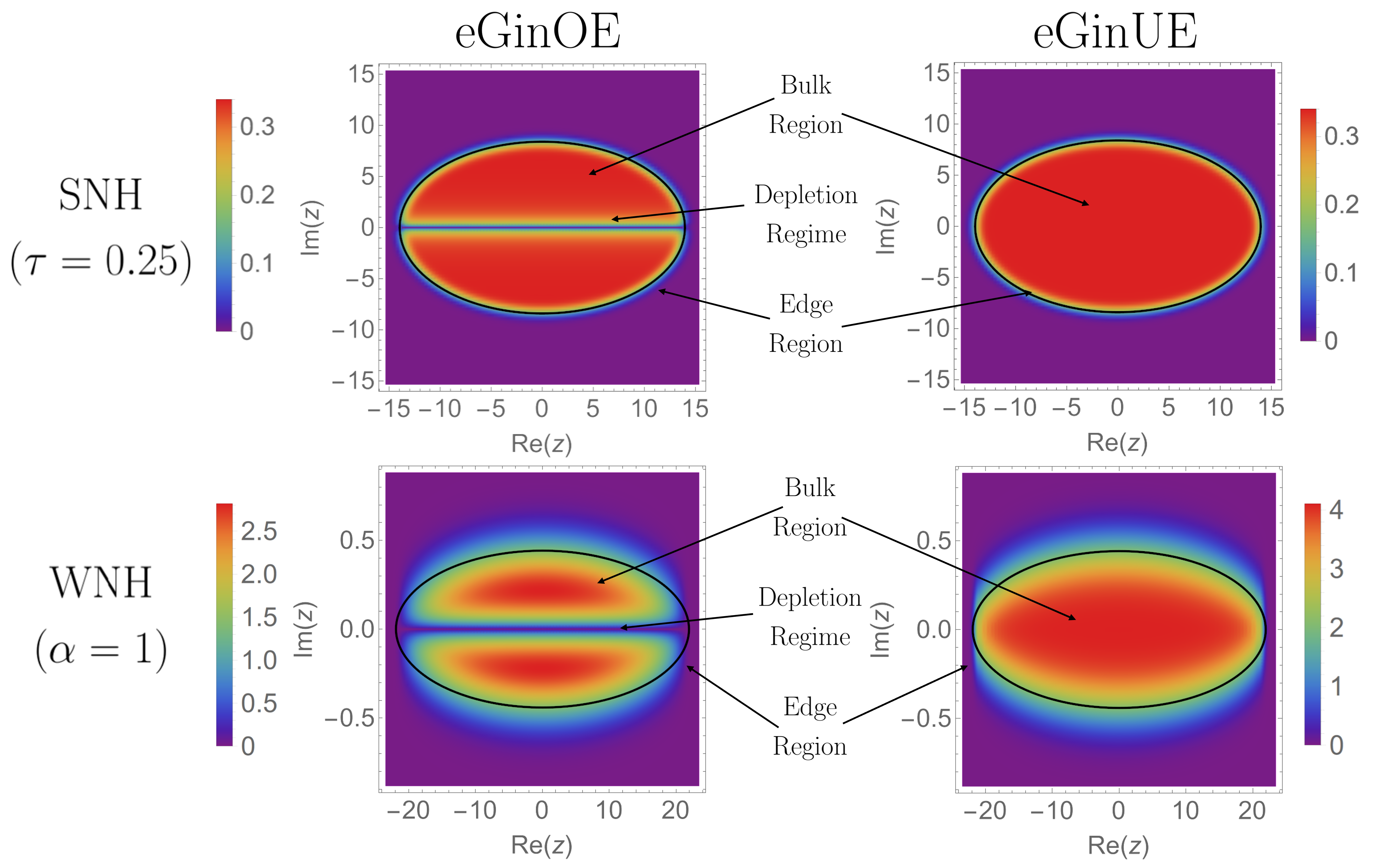}
    
    \caption{\small Large-$N$ density of complex eigenvalues in the eGinOE (left) and eGinUE (right) in the strong (top) and weak (bottom) non-Hermiticity regimes, with relevant scaling regions labelled. Strong non-Hermiticity plots are shown for $\tau = 0.25$ and weak non-Hermiticity plots are shown for $\alpha = 1$. In all plots, the density is depicted on a rainbow scale and the solid black ellipse has semi-major and semi-minor axes $\sqrt{N}(1+\tau)$ and $\sqrt{N}(1-\tau)$ respectively. Each heatmap is plotted using the finite-$N$ expression for the density of complex eigenvalues in the associated ensemble when $N=125$. This diagram was originally presented in \cite{CFW24}. }
    \label{fig:Heatmaps}
\end{figure}

It is well-known that, for large-$N$, the complex eigenvalues of both ensembles fill the interior of an ellipse in the complex plane, with semi-major and semi-minor axes given by $\sqrt{N}(1+\tau)$ and $\sqrt{N}(1-\tau)$ respectively \cite{Girko2,SCSS}. In the case of the eGinUE, this ellipse features two very distinct non-zero regimes of eigenvalue density for both SNH and WNH, a spectral bulk and an edge region. On the other hand, the eGinOE features both of these regions as well as an additional region of eigenvalue depletion close to the real line, due to the presence of a macroscopic number of real eigenvalues. These regimes are depicted in both ensembles for strong and weak non-Hermiticity in Figure \ref{fig:Heatmaps}. 

The other main object of interest in this work is the mean self-overlap between left and right eigenvectors. This object was recently calculated at finite-$N$ and for $\tau \in[0,1)$ in \cite{CFW24} for both the eGinUE and eGinOE. In the eGinUE it was found that
\begin{equation}
    \mathcal{O}^{(\eGinUE,c)}_{N}(z) = \rho^{(\textup{eGinUE},c)}_{N}(z) + (1-\tau^2) \bigg[ \,  \rho^{(\textup{eGinUE},c)}_{N-1}(z) + (N-2) \, \rho^{(\textup{eGinUE},c)}_{N-2}(z) - R_{N-3} \, \bigg] \ ,
    \label{eq:O_N_eGinUE}
\end{equation}
where $\rho^{(\textup{eGinUE},c)}_{N}(z)$ is defined in Eq. \eqref{eq:rho_N_eGinUE} and 
\be\label{eq:R_N}
\begin{split}
    R_N &
    \equiv \frac{1}{\pi } \ \frac{1}{\sqrt{1-\tau^2}}  \ \exp \left\{ -\frac{\vert z \vert^2 - \tau \, \RE (z^2)}{1-\tau^2} \right\} \sum_{k=0}^{N} \ \frac{k \ \tau^{k}}{k!} \, \HE_k\left(\frac{\bar{z}}{\sqrt{\tau}}\right) \HE_{k}\left( \frac{z}{\sqrt{\tau}} \right)  \ .
\end{split}
\ee
Correspondingly, in the eGinOE it was found that
\begin{align}
    \mathcal{O}^{(\eGinOE,c)}_{N}(z) &= \frac{1}{\pi} \ \sqrt{\frac{1-\tau}{1+\tau}} e^{ -\frac{\RE(z)^2}{1+\tau} -\frac{\IM(z)^2}{1-\tau} }   \bigg[ 1 + \sqrt{\frac{\pi(1-\tau^2)}{2}} e^{ \frac{2 \IM(z)^2}{1-\tau^2} } \, \frac{1}{2\vert \IM(z) \vert} \ \textup{erfc}\left( \sqrt{\frac{2 }{1-\tau^2}} \ \vert \IM(z) \vert \right) \bigg] \nonumber \\
    &\times \bigg[   P_{N-2} + (1-\tau^2)\bigg( P_{N-3} + (N-3) P_{N-4} - T_{N-4} \bigg) \bigg]  \ ,
    \label{eq:O_N_eGinOE}
\end{align}
where $P_N$ is defined in Eq. \eqref{eq:P_N} and 
\begin{equation}
    T_N \equiv \frac{1}{\bar{z}-z} \sum_{k=0}^{N} \ \frac{k \ \tau^{k+\frac{1}{2}}}{k!} \bigg[ \HE_{k+1}\left(\frac{\bar{z}}{\sqrt{\tau}}\right) \HE_{k}\left( \frac{z}{\sqrt{\tau}} \right) - \HE_{k+1}\left( \frac{z}{\sqrt{\tau}} \right) \HE_{k}\left(\frac{\bar{z}}{\sqrt{\tau}}\right) \bigg] \ .
    \label{eq:T_N}
\end{equation}

\begin{rem}
    This manuscript will not discuss the density of complex eigenvalues or the mean self-overlap in the bulk or depletion regime, for neither SNH nor WNH, but will focus solely on the edge region. Results for the bulk region of the eGinOE and eGinUE and the depletion regime of the eGinOE can be found in \cite[Section 2]{CFW24}. One can also find more information on the correlation structure and methods of sampling matrices in the eGinUE and eGinOE within \cite{CFW24}. Additionally, for a review of other results available in the eGinUE and eGinOE one can consult \cite{BF1} and \cite{BF2} respectively.
\end{rem}

\noindent
In the work by Lee and Riser \cite{LR}, the authors consider the density of eigenvalues at the edge of the eGinUE in the strong non-Hermiticity regime. Here, they derive the leading order term in $N$ and the first two subleading correctional terms. In their setting, they find that the leading order term is independent of the curvature of the ellipse and solely depends on the perpendicular distance from the ellipse edge. Additionally, the density of complex eigenvalues at the edge of the eGinUE at weak non-Hermiticity can be found within the work by Garci\'{a}-Garci\'{a} et. al. \cite{GGNV}, who derive the correlation kernel of the corresponding eigenvalue point process in this limit, from which the density follows straightforwardly. These results, as well as corresponding results by Akemann, Bender \& Phillips in the eGinOE \cite{AB, AP}, provide useful points of comparison and so we state their findings as part of the following section. 

\subsection{Statement of Main Results}
\label{sec:Main_Results}

In this section, we outline the main results and definitions that are central to this paper. We begin by providing results relevant to the strong non-Hermiticity regime and then move onto the weak non-Hermiticity regime. At the beginning of each section, we provide a definition of the edge in each specific regime of $\tau$. 

\subsubsection{Strong non-Hermiticity Regime}
\label{sec:SNH_edge_results}

\begin{figure}[h]
    \centering
    \includegraphics[scale = 0.3]{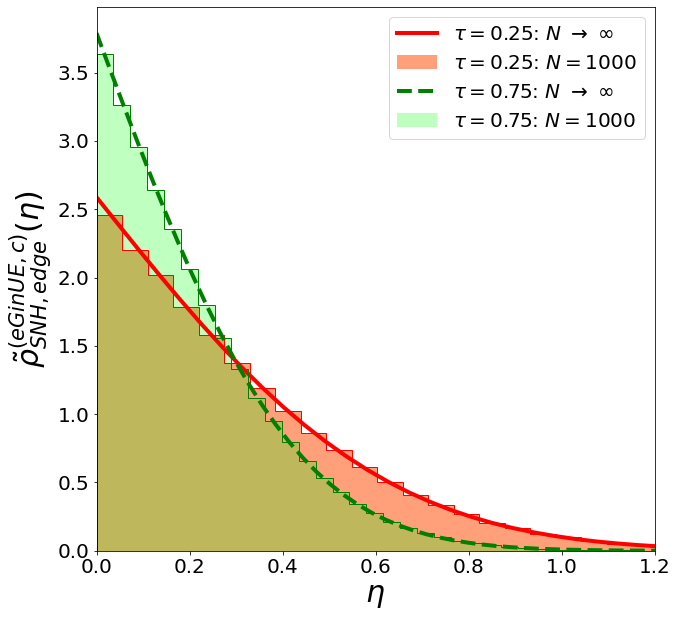}
    \hspace{1cm}
    \includegraphics[scale = 0.3]{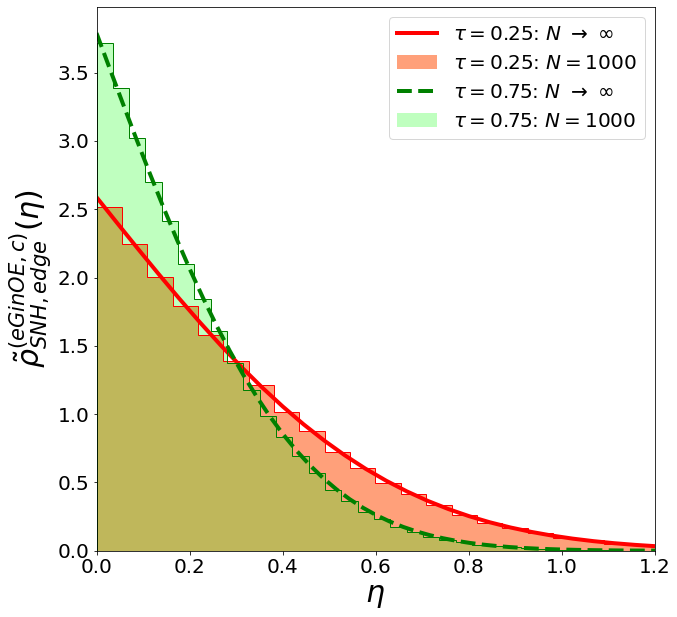}
    \caption{\small Normalised density of complex eigenvalues at the edge of the eGinUE (left) and eGinOE (right) at strong non-Hermiticity as a function of $\eta$, for $\tau =0.25$ (red) and $\tau = 0.75$ (green). In both plots the solid (dotted) line is plotted using Eq. \eqref{eq:rho_tilde} with $\tau = 0.25$ (0.75). Each histogram is generated by retaining eigenvalues with $\eta > 0$, from a sample containing $O(10^7)$ eigenvalues of eGinUE/eGinOE matrices of size $N=1000$. In the eGinOE plot, eigenvalues with $|\sin(\theta)| \sim O(1)$ are not included, so as to avoid the depletion regime. }
    \label{fig:rho_SNH_edge}
\end{figure}

In order to state our main results for the eGinUE and eGinOE at strong non-Hermiticity, we first carefully define the scaling necessary to describe the edge region of the droplet. A point on the boundary of the elliptic droplet is defined as
\begin{equation}
    z_0 = \sqrt{N}(1 + \tau) \cos(\theta) + i \sqrt{N}(1 - \tau) \sin(\theta) \ ,  
    \label{eq:z_0}
\end{equation}
where, in the eGinUE, we have $0 \leq \theta < 2 \pi$. However, in order to avoid the edge of the depletion regime in the eGinOE, we only consider points where $|\sin(\theta)| \sim O(1)$. The unit normal vector from the point $z_0$ can be presented in the form
\begin{equation}
    \widehat{\bm n} = \frac{1}{\sqrt{1 + \tau^2 - 2 \tau \cos(2\theta)}} \left( \begin{matrix} (1 - \tau) \cos(\theta) \\ (1 + \tau) \sin(\theta) \end{matrix} \right) \ ,
\end{equation}
which we utilise to define another point in the vicinity of the edge, denoted as $z$. Specifically, we use a small perturbation, $\eta \, \widehat{\bm n}$, from $z_0$, such that we can write the real and imaginary parts of $z$ as 
\begin{equation}
    \begin{aligned}
        &\RE(z) = \sqrt{N}(1 + \tau) \cos(\theta) + \frac{ \eta  (1 - \tau) \cos(\theta) }{\sqrt{ 1 + \tau^2 - 2 \tau \cos(2 \theta)}} = \sqrt{N} x_e + w_x \\	
        &\IM(z) = \sqrt{N}(1 - \tau) \sin(\theta) + \frac{ \eta  (1 + \tau) \sin(\theta) }{\sqrt{ 1 + \tau^2 - 2 \tau \cos(2 \theta)}} = \sqrt{N} y_e + w_y \ .
        \label{eq:edge_xy_params} 
    \end{aligned}
\end{equation} 
In order to obtain the density of complex eigenvalues for the edge of the eGinUE and eGinOE at SNH in this setting, one can utilise the above definition of $z$ in the corresponding finite-$N$ density equation, for a fixed $\tau \in [0,1)$ and then take $N \to \infty$. This yields the following universal result:
\begin{prop}\label{prop:rho_eGinOE_SNH_edge}
    Let $X$ be an $N \times N$ matrix drawn from the eGinUE via Eq. \eqref{eq:jpdf_eGinUE}. Let $z$ be a complex eigenvalue of $X$, scaled in the strong non-Hermiticity limit, where $\tau \in [0,1)$ is fixed and $z$ obeys Eq. \eqref{eq:edge_xy_params}. In this limit, as $N$ becomes large, the mean density of complex eigenvalues reads
    \begin{align}
        \rho_{\textup{SNH,edge}}^{\textup{(eGinUE,c)}}(\eta) = \lim_{N\to \infty} \rho_N^{\textup{(eGinUE,c)}}\Big( \big(\sqrt{N} x_e + w_x\big) + i \big(\sqrt{N} y_e + w_y\big) \Big) = \frac{1}{2\pi(1 - \tau^2)} \textup{erfc}\left( \sqrt{\frac{2}{1 - \tau^2}} \, \eta \right) \ .
        \label{eq:rho_eGinUE_SNH_edge}
    \end{align}
    Correspondingly if $X$ is drawn from the eGinOE via Eq. \eqref{eq:jpdf_eGinOE}, with $|\sin(\theta)|\sim O(1)$, the density reads 
    \begin{align}
        \rho_{\textup{SNH,edge}}^{\textup{(eGinOE,c)}}(\eta) = \lim_{N\to \infty} \rho_N^{\textup{(eGinOE,c)}}\Big( \big(\sqrt{N} x_e + w_x\big) + i \big(\sqrt{N} y_e + w_y\big) \Big) = \frac{1}{2\pi(1 - \tau^2)} \textup{erfc}\left( \sqrt{\frac{2}{1 - \tau^2}} \, \eta \right) \ .
        \label{eq:rho_eGinOE_SNH_edge}
    \end{align}
\end{prop}
\noindent
One notes that these expressions are independent of the angle $\theta$ and only depend on $\eta$. The above result for the eGinUE coincides with the first order of the expansion found by Lee and Riser \cite{LR}, see alternatively \cite{AP}. Whereas the statement for the case of the eGinOE, aligns with the result of Akemann and Phillips \cite{AP}, derived from the corresponding correlation kernel in \cite[Eq. (5.81)]{AP}, using orthogonal polynomial methods. For the convenience of the reader, in Appendix \ref{app:rho_eGinOE_SNH}, we outline the derivation of Eq. \eqref{eq:rho_eGinOE_SNH_edge} from the finite-$N$ result in Eq. \eqref{eq:rho_N_eGinOE}. The corresponding derivation of Eq. \eqref{eq:rho_eGinUE_SNH_edge} from Eq. \eqref{eq:rho_N_eGinUE} follows a similar and slightly simpler method. In Figure \ref{fig:rho_SNH_edge} we plot the normalised density of complex eigenvalues with $\eta>0$,
\begin{equation}
    \widetilde{\rho}^{\, \text{(ensemble)}}_{\, \text{SNH,edge}}(\eta) = \frac{\rho_{\text{SNH, edge}}^{\text{(ensemble)}}(\eta)}{\int_0^\infty \rho_{\text{SNH, edge}}^{\text{(ensemble)}}(\eta') \, d \eta'} \ ,
    \label{eq:rho_tilde}
\end{equation}
for both the eGinUE and eGinOE, with $\tau = 0.25$ and $\tau=0.75$. The universality of this density is evident from the Figure as the results for the eGinUE and eGinOE are nearly identical.\\

\noindent
One can then extend this analysis so as to consider the mean self-overlap of eigenvectors at the edge, in both the eGinUE and eGinOE. This is achieved by starting from the finite-$N$ results in Eqs. \eqref{eq:O_N_eGinUE} and \eqref{eq:O_N_eGinOE} and choosing the edge to be defined as in Eq. \eqref{eq:edge_xy_params}. Keeping $\tau \in [0,1)$ fixed and taking $N \to \infty$, one is able to obtain the leading order behaviour of the mean self-overlap at the edge in the eGinUE and eGinOE.

\begin{thm} \label{thm:O_eGinOUE}
    Let $X$ be an $N \times N$ matrix drawn from the eGinUE via Eq. \eqref{eq:jpdf_eGinUE}. Let $z$ be a complex eigenvalue of $X$, scaled in the strong non-Hermiticity limit, where $\tau \in [0,1)$ is fixed and $z$ obeys Eq. \eqref{eq:edge_xy_params}. In this limit, as $N$ becomes large, the mean self-overlap of the associated left and right eigenvectors reads
    \begin{equation}
    \begin{aligned}
        \mathcal{O}_{\textup{SNH,edge}}^{\textup{(eGinUE,c)}}(\eta, \theta) =& \lim_{N \to \infty} \frac{1}{\sqrt{N}} \mathcal{O}_N^{\textup{(eGinUE,c)}} \Big( \big(\sqrt{N} x_e + w_x\big) + i \big(\sqrt{N} y_e + w_y\big) \Big) \\
        =& \frac{1}{\pi} \sqrt{\frac{1 + \tau^2 - 2 \tau \cos(2 \theta)}{1 - \tau^2}} \left[ \frac{1}{\sqrt{2 \pi}} \exp{  \frac{-2\eta^2}{1 - \tau^2} }  - \frac{\eta}{\sqrt{{1 - \tau^2}}} \textup{erfc}\left( \sqrt{\frac{2}{1 - \tau^2}} \eta \right)\right] \ . 
        \label{eq:O_SNH_edge_eGinUE}
    \end{aligned}
    \end{equation}
    Correspondingly, if one considers the same quantity in the eGinOE, with the additional constraint in Eq. \eqref{eq:edge_xy_params} that $|\sin(\theta)| \sim O(1)$, one finds that:
    \begin{equation}
    \begin{aligned}
        \mathcal{O}_{\textup{SNH,edge}}^{\textup{(eGinOE,c)}}(\eta, \theta) =& \lim_{N \to \infty} \frac{1}{\sqrt{N}} \mathcal{O}_N^{\textup{(eGinOE,c)}} \Big( \big(\sqrt{N} x_e + w_x\big) + i \big(\sqrt{N} y_e + w_y\big) \Big)  \\
        =& \frac{1}{\pi} \sqrt{\frac{1 + \tau^2 - 2 \tau \cos(2 \theta)}{1 - \tau^2}} \left[ \frac{1}{\sqrt{2 \pi}} \exp{ \frac{-2\eta^2}{1 - \tau^2} } - \frac{\eta}{\sqrt{{1 - \tau^2}}} \textup{erfc}\left( \sqrt{\frac{2}{1 - \tau^2}} \eta \right)\right] \label{eq:O_SNH_edge_eGinOE} \ .
    \end{aligned}
    \end{equation}
\end{thm}
\noindent
This directly leads to the following Corollary:
\begin{cor}
     In the strong non-Hermiticity limit, for both the eGinUE and eGinOE, the mean weighted conditional self-overlap of eigenvectors associated with a complex eigenvalue $z$, scaled via Eq. \eqref{eq:edge_xy_params}, reads:
    \begin{equation}
    \begin{aligned}
        \widetilde{\mathbb{E}}^{\textup{(eGin,c)}}_{\textup{SNH,edge}}(\Tilde{\eta}) =& \frac{1}{2\sqrt{(1-\tau^2)(1 + \tau^2 - 2 \tau \cos(2 \theta))}} \frac{\mathcal{O}_{\textup{SNH,edge}}^{\textup{(eGin,c)}}(\eta, \theta)}{ \rho_{\textup{SNH,edge}}^{\textup{(eGin,c)}}(\eta)}\\
        =& \frac{1}{\textup{erfc}(\sqrt{2} \Tilde{\eta})}\left[ \frac{1}{\sqrt{2 \pi}} \exp{ - 2 \widetilde{\eta}^{\, 2}} - \widetilde{\eta} \, \textup{erfc}(\sqrt{2} \widetilde{\eta}) \right]  \ .
        \label{eq:Etilde_eGinOUE}
    \end{aligned}
    \end{equation}
    Here we have used eGin as a collective shorthand to denote both the eGinOE and eGinUE and $\Tilde{\eta} = \eta/\sqrt{1 - \tau^2}$.
\end{cor}

\begin{figure}[h]
    \centering
    \includegraphics[scale=0.3]{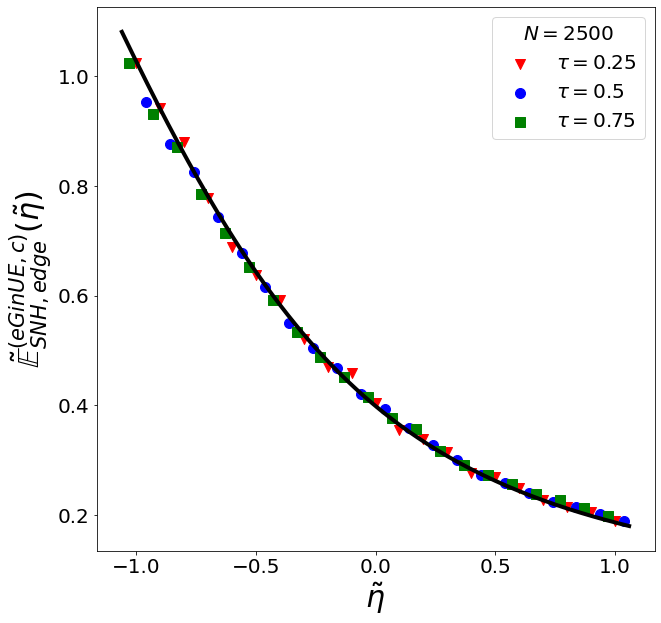}
    \hspace{1cm}
    \includegraphics[scale=0.3]{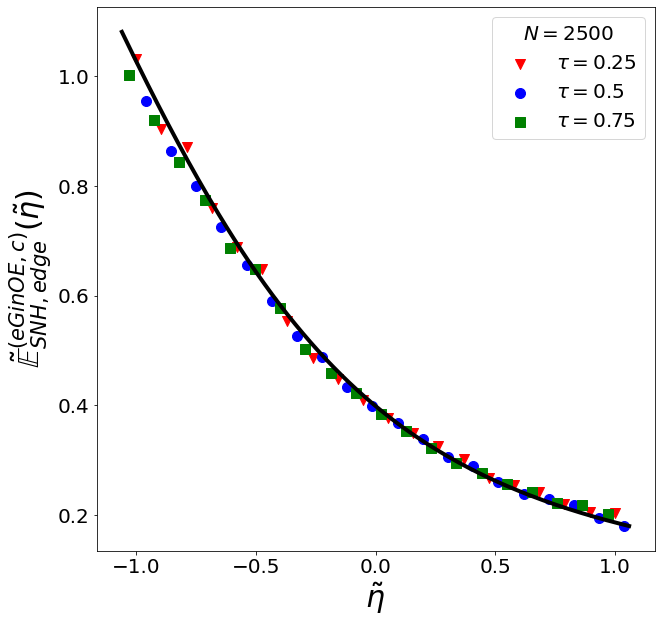}
    \caption{\small Mean weighted conditional self-overlap of eigenvectors associated with complex eigenvalues at the elliptic edge in the eGinUE (left) and eGinOE (right) at strong non-Hermiticity, for a range of $\tau$ (coloured markers). The solid black line in both plots is the universal prediction for the mean weighted conditional self-overlap from Eq. \eqref{eq:Etilde_eGinOUE}. Each coloured marker is a numerical average of self-overlaps associated with complex eigenvalues within $\pm 1/\sqrt{N}$ of the specified $\eta$, taken from a set containing $O(10^6)$ eigenvalues of eGinUE and eGinOE matrices of size $N=2500$. Note that, in order to avoid the depletion regime in the eGinOE, we do not consider eigenvalues with imaginary component smaller than one. }
    \label{fig:E_eGinOUE_SNH_edge}
\end{figure}

\noindent
These results establish universality of the leading order behaviour of the mean self-overlap at the edge for the eGinUE and eGinOE, in the strong non-Hermiticity regime. Note that, in contrast to the mean density, the limiting mean self-overlap in both ensembles is dependent on both $\eta$ and $\theta$. In Figure \ref{fig:E_eGinOUE_SNH_edge}, we compare these findings for the edge in the eGinUE and eGinOE at strong non-Hermiticity to direct numerical simulation of the mean weighted conditional self-overlap. These plots demonstrate a good agreement between theory and simulations when $N=2500$, which is a much larger value of $N$ than was required for previous results pertaining to the spectral bulk \cite{CFW24}. This is due to the fact that the leading order asymptotic term in the mean self-overlap is proportional to $\sqrt{N}$ at the edge, as opposed to $N$ in the bulk. 

\begin{rem}
    Setting $\tau = 0$ in Eqs. \eqref{eq:rho_eGinUE_SNH_edge}, \eqref{eq:rho_eGinOE_SNH_edge}, \eqref{eq:O_SNH_edge_eGinUE} and \eqref{eq:O_SNH_edge_eGinOE} accurately reproduces results pertaining to the edge of the classical Ginibre ensembles, GinUE and GinOE, which are independent of $\theta$, as outlined in \cite{WCF23}. Therefore, one might have suspected similar angular independence to be observed for the mean self-overlap in the elliptic Ginibre ensembles. However, upon further consideration, the angular dependence in Eqs. \eqref{eq:O_SNH_edge_eGinUE} and \eqref{eq:O_SNH_edge_eGinOE} can be reconciled by noting that in \cite{CFW24,WCF23} the self-overlap is systematically reduced as eigenvalues move further away from the origin. Hence, in an elliptic ensemble at fixed $\eta$, it is unsurprising that there is a non-trivial dependence on $\theta$. 
\end{rem}

\subsubsection{Weak non-Hermiticity Regime} \label{sec:WNH_edge_results}

In the limit of weak non-Hermiticity, the correlation parameter $\tau$ approaches unity asymptotically as $N \to \infty$, hence the droplet is reduced to a thin strip, close to the real line. We define this regime such that 
\begin{equation}
    \tau = 1 - \frac{\kappa}{N} \, , \hspace{1cm} \text{and} \hspace{1cm} \kappa = \frac{(\pi \alpha)^2}{2} \ ,
    \label{eq:tau_WNH}
\end{equation}
where the choice of $\kappa$ is to allow one to make connections with existing literature results \cite{CFW24,ACV}. We choose to parameterise the edge in the same way as in \cite{GGNV}, such that
\begin{equation}
    z = 2 \sqrt{N} + \frac{x}{N^{1/6}} + i \frac{y}{N^{1/2}} \ .
    \label{eq:z_edge_WNH}
\end{equation}
Strictly speaking, this should be referred to as the rightmost edge of the ellipse, however, due to symmetry arguments, these results also hold at the left-most edge, where $\RE(z) =- 2\sqrt{N} - xN^{-1/6}$. For simplicity we focus on $y>0$ henceforth, note however that results can also be extrapolated to negative $y$ via simple symmetry arguments. In the above edge scaling at weak non-Hermiticity, it is known that, in the eGinUE, the limiting density of complex eigenvalues reads \cite{GGNV}
\begin{align}
    \rho_{\textup{WNH,edge}}^{\textup{(eGinUE,c)}}(x,y) =& \lim_{N \to \infty} \frac{1}{N^{2/3}} \rho_N^{\textup{(eGinUE,c)}} \bigg( z = 2 \sqrt{N} + \frac{x}{N^{1/6}} + i \frac{y}{N^{1/2}} \bigg)
    = \frac{1}{\sqrt{\pi \kappa}} e^{-\frac{y^2}{\kappa}} \int_0^\infty dt \, \textup{Ai} (x+t)^2 \ , \label{eq:rho_eGinUE_WNH_edge} 
\end{align}
where we have made use of the well-known Airy function, defined as
\begin{equation}
    \textup{Ai}(z) = \frac{1}{2\pi} \int_{-\infty}^\infty dt \, \exp{i\left( \frac{t^3}{3} + zt \right)} = \frac{1}{2 \pi i} \int_{\cont} dp\, \exp{\frac{p^3}{3} - zp} \ ,
    \label{eq:Airy_func}
\end{equation}
such that ${\cont}$ is a contour that starts at $\infty$ with an argument of $-\pi/3$ and ends at $\infty$ with an argument of $\pi/3$. This result can also be deduced from \cite[Prop. 4]{AB}, using a rescaling of the imaginary parts and taking the proper limits. The statement below provides the counterpart of the density in Eq. \eqref{eq:rho_eGinUE_WNH_edge}, for the case of the eGinOE, in the scaling defined in Eq. \eqref{eq:z_edge_WNH}.

\begin{prop} \label{prop:rho_eGinOE_WNH_edge}
    Let $X$ be an $N \times N$ matrix drawn from the eGinOE, via Eq. \eqref{eq:jpdf_eGinOE}. Let $z$ be a complex eigenvalue of $X$, scaled in the weak non-Hermiticity limit, where $\tau = 1 - \kappa/N$ and $z$ obeys Eq. \eqref{eq:z_edge_WNH}. In this limit, as $N$ becomes large, the mean density of complex eigenvalues reads  
    \begin{equation}
    \begin{aligned}
        \rho_{\textup{WNH,edge}}^{\textup{(eGinOE,c)}}(x,y) =& \lim_{N \to \infty} \rho_{N}^{\textup{(eGinOE,c)}}\bigg( 2 \sqrt{N} + \frac{x}{N^{1/6}} + i \frac{y}{N^{1/2}}  \bigg) \label{eq:rho_eGinOE_WNH_edge} \\
        =& \, y \, \textup{erfc}\left( \frac{y}{\sqrt{\kappa}} \right) \int_0^\infty dt \bigg[ \textup{Ai}'(x+t)^2 - \textup{Ai}(x+t) \, \textup{Ai}''(x+t) \bigg] \ .
    \end{aligned}
    \end{equation}
    This is given in terms of $\textup{Ai}'(w)$ and $\textup{Ai}''(w)$, which denote the first and second derivatives of the Airy function, defined in Eq. \eqref{eq:Airy_func}, evaluated at the point $w$. 
\end{prop}

\begin{figure}[h]
    \centering
    \includegraphics[scale = 0.3]{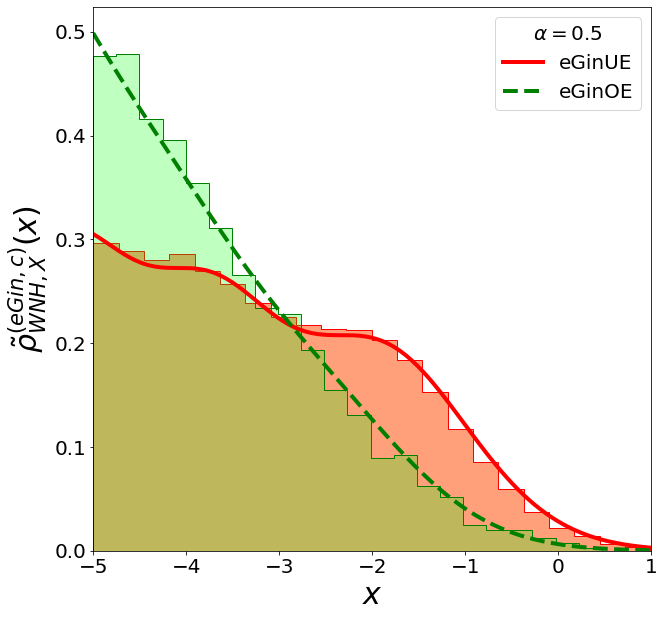}
    \hspace{1cm}
    \includegraphics[scale = 0.3]{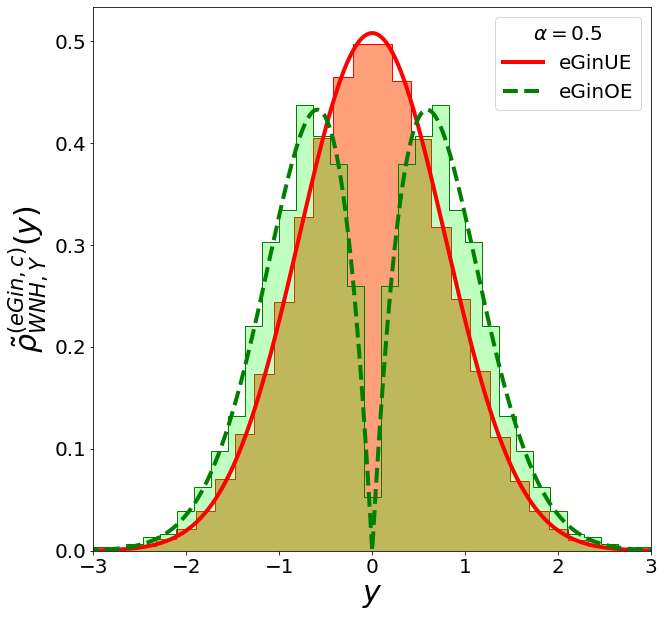}
    \caption{\small Normalised density of scaled real (left) and imaginary (right) components of complex eigenvalues at the edge of the elliptic droplet for eGinOE (green) and eGinUE (red) matrices. This is shown for the weak non-Hermiticity regime of $\tau$, with fixed $\alpha = 0.5$. Each plot contains normalised histograms of eigenvalues obtained through numerical diagonalisation of $O(10^7)$ eGinUE and eGinOE matrices of size $N=1000$. Solid (dashed) lines are generated from the theoretical density of complex eigenvalues for the eGinUE (eGinOE) and normalised according to Eq. \eqref{eq:rho_tilde_WNH}.}
    \label{fig:rho_edge_WNH}
\end{figure}

\noindent
In \cite{AB, AP} a different scaling of real and imaginary parts is used such that they are scaled with the same power of $N$. Therein, the authors find the correlation for kernel both the eGinUE and eGinOE to leading order at WNH. To reconnect with the statement above, a proper rescaling as well as integration by parts is needed to derive Eq. \eqref{eq:rho_eGinUE_WNH_edge} and Eq. \eqref{eq:rho_eGinOE_WNH_edge}, starting from their respective expressions of the correlation kernels. Note that both densities given in Eqs. \eqref{eq:rho_eGinUE_WNH_edge} and \eqref{eq:rho_eGinOE_WNH_edge} can have their $x$ and $y$ dependencies factorised such that $\rho_{\text{WNH,edge}}^{\textup{(eGin,c)}}(x,y) = \rho^{\textup{(eGin,c)}}_{\text{WNH},X}(x) \rho^{\textup{(eGin,c)}}_{\text{WNH},Y}(y)$. This indicates that, at WNH, the leading order behaviour of the real and imaginary components of edge eigenvalues are independent variables. Therefore, we introduce the normalised density of the real and imaginary components, through the use of 
\begin{equation}
    \widetilde{\rho}^{\textup{(eGin,c)}}_{\text{WNH},X}(x) = \frac{\rho^{\textup{(eGin,c)}}_{\text{WNH},X}(x)}{\int_{S_R}\rho^{\textup{(eGin,c)}}_{\text{WNH},X}(x') dx'} \hspace{0.5cm} \text{and} \hspace{0.5cm} \widetilde{\rho}^{\textup{(eGin,c)}}_{\text{WNH},Y}(y) = \frac{\rho^{\textup{(eGin,c)}}_{\text{WNH},Y}(y)}{\int_{-\infty}^\infty\rho^{\textup{(eGin,c)}}_{\text{WNH},Y}(y') dy'} \ ,
    \label{eq:rho_tilde_WNH}
\end{equation}
where $S_R$ is some subset of the real line. In Figure \ref{fig:rho_edge_WNH}, we plot the above distributions and compare them to results from direct numerical simulation of large eGinUE and eGinOE matrices, choosing to use $S_R = [-5,1]$.\\

\noindent
We now consider the mean self-overlap at the edge for both the eGinUE and eGinOE at WNH. In contrast to what was found in the case of SNH, the results here do not turn out to be universal across the two ensembles.

\begin{rem}
     At weak non-Hermiticity in the eGinUE, it can be seen that the mean self-overlap of  eigenvectors associated with complex edge eigenvalues, asymptotically approaches the density, i.e. $\mathcal{O}(z) \approx \rho(z)$ as $N \to \infty$. As a consequence of this, the mean conditional self-overlap goes to unity, i.e, $\mathbb{E} \to 1$. For this reason, in the eGinUE, we make the choice to study the mean \emph{shifted} conditional self-overlap, $\mathcal{S} = \mathbb{E} - 1$, so as to obtain a non-trivial result to compare to numerical data.
\end{rem}

\begin{thm} \label{thm:S_eGinUE_WNH_edge}
    Let $X$ be an $N \times N$ matrix drawn from the eGinUE via Eq. \eqref{eq:jpdf_eGinUE}. Let $z$ be a complex eigenvalue of $X$, scaled in the weak non-Hermiticity limit, where $\tau = 1 - \kappa/N$ and $z$ obeys Eq. \eqref{eq:z_edge_WNH}. In this limit, as $N$ becomes large, the mean shifted conditional self-overlap of the associated left and right eigenvectors reads
    \begin{equation}
    \begin{aligned}
        \mathcal{S}^{\textup{(eGinUE,c)}}_{\textup{WNH, edge}}(x) =& \lim_{N \to \infty} N^{2/3} \bigg( \mathbb{E}^{\textup{(eGinUE,c)}}_{N} \left( z = 2 \sqrt{N} + \frac{x}{N^{1/6}} + i \frac{y}{N^{1/2}} \right) - 1 \bigg) \label{eq:S_eGinUE_WNH_edge} = \frac{2 \kappa \int_0^\infty dt \, t \, \textup{Ai}(x+t)^2}{\int_0^\infty dt  \, \textup{Ai}(x+t)^2} \ ,
    \end{aligned}
    \end{equation}
    where $\textup{Ai}(\cdot)$ denotes the Airy function, defined in Eq. \eqref{eq:Airy_func}.
\end{thm}

\noindent
When considering the mean self-overlap at the edge of the eGinOE at WNH, one finds contrasting behaviour to the eGinUE in the same limit. Namely, due to the presence of the depletion regime, the mean self-overlap does not approach the density as $N$ becomes large. Hence, in this case, it does make sense to study the conventional mean conditional self-overlap, with the result given by the next Theorem:

\begin{thm}\label{thm:E_eGinOE_WNH_edge}
Let $X$ be an $N \times N$ matrix drawn from the eGinOE via Eq. \eqref{eq:jpdf_eGinOE}. Let $z$ be a complex eigenvalue of $X$, scaled in the weak non-Hermiticity limit, where $\tau = 1 - \kappa/N$ and $z$ obeys Eq. \eqref{eq:z_edge_WNH}. In this limit, as $N$ becomes large, the mean conditional self-overlap of the associated left and right eigenvectors reads
\begin{equation}
\begin{aligned}
    \mathbb{E}_{\textup{WNH,edge}}^{\textup{(eGinOE,c)}}(y) =& \lim_{N \to \infty} \mathbb{E}^{\textup{(eGinOE,c)}}_{N} \bigg( z = 2 \sqrt{N} + \frac{x}{N^{1/6}} + i \frac{y}{N^{1/2}} \bigg)  \\
    =&\frac{1}{y \, \textup{erfc}\left( \frac{y}{\sqrt{\kappa}} \right)} \bigg[ \sqrt{\frac{\kappa}{\pi}} \exp{\frac{-y^2}{\kappa}} + \frac{\kappa}{2y} \textup{erfc} \left( \frac{y}{\sqrt{\kappa}} \right)\bigg] \ .
    \label{eq:E_eGinOE_WNH_edge}
\end{aligned}
\end{equation}
\end{thm}

\begin{figure}[h]
    \centering
    \includegraphics[scale = 0.3]{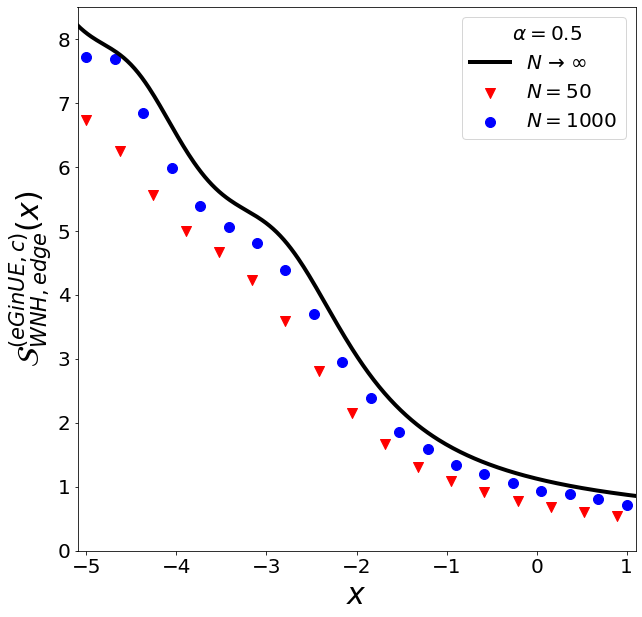}
    \hspace{1cm}
    \includegraphics[scale = 0.3]{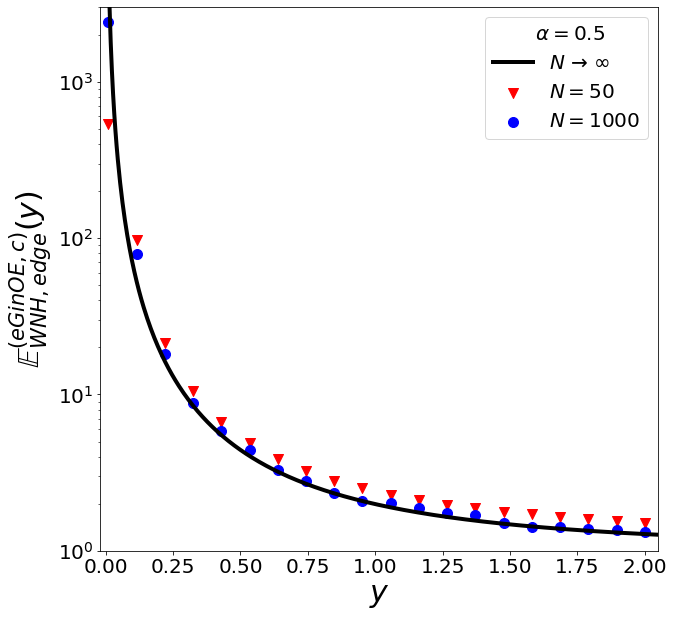}
    \caption{\small Conditional self-overlap of eigenvectors associated with complex eigenvalues at the edge of elliptic Ginibre ensembles at weak non-Hermiticity, with fixed $\alpha = 0.5$ for large finite-$N$. Left: mean shifted conditional self-overlap in the eGinUE, $\mathcal{S}^{\text{(eGinUE,c)}}_{\text{WNH,edge}}(x)$, as a function of $x$. Right: mean conditional self-overlap in the eGinOE, $\mathbb{E}^{\text{(eGinOE,c)}}_{\text{WNH,edge}}(y)$, as a function of $y$. The solid black line is plotted according to Eqs. \eqref{eq:S_eGinUE_WNH_edge} and \eqref{eq:E_eGinOE_WNH_edge} in the left and right hand plots respectively. Each coloured marker is a numerical average of self-overlaps observed in simulation of eGinUE/eGinOE matrices, taken from a sample containing $O(10^7)$ eigenvalues. In simulation, self-overlaps associated with a fixed $x$ are obtained by retaining all $x$ within $\pm 0.1$ of the desired $x$, for all $y$. The opposite is true for a fixed $y$. }
    \label{fig:eGinOUE_WNH_edge}
\end{figure}

\noindent
In Figure \ref{fig:eGinOUE_WNH_edge}, we plot the mean shifted conditional self-overlap for the eGinUE and the mean conditional self-overlap for the eGinOE and compare to direct numerical simulation of the self-overlap of left and right eigenvectors associated with complex eigenvalues at the edge. In both plots, one finds a good agreement with theoretical predictions at large-$N$, with minor discrepancies attributed to finite-$N$ effects. This is particularly relevant to the results pertaining to the eGinUE (left-hand plot).

\begin{rem}
    The results outlined in this section show that, when comparing the eGinUE and eGinOE, the density of complex eigenvalues and the mean self-overlap are not universal quantities at the edge, in the limit of weak non-Hermiticity. This phenomenon is not unexpected and follows from the fact that the eGinOE has a macroscopic fraction of real eigenvalues and a region of eigenvalue depletion close to the real line, whereas all the eigenvalues are complex in the case of the eGinUE. 
\end{rem}

\subsection{Discussion of Open Problems}
\label{sec:open_probs}

Despite many recent advances in the study of eigenvalues and eigenvectors in the Ginibre ensembles, some questions still remain open. Universality of the density at the edge is established across the elliptic Ginibre ensembles at SNH, with results for the eGinSE available in \cite{AP, ByunEbke, ABK, AKMP}. In contrast, there is no such universality to be extended in the realm of weak non-Hermiticity, therefore, the edge density of complex eigenvalues is very much an open question in the WNH regime of the eGinSE. 

Similar questions remain as to the universality of the mean self-overlap. One would conjecture that the results of Theorem \ref{thm:O_eGinOUE} could be extended to the eGinSE at SNH, due to the expected universality. On the other hand, at WNH, the corresponding results in the eGinSE are entirely unknown. Furthermore, in comparison to the eGinOE and eGinUE, results for the mean self-overlap are more scarce in the eGinSE, as questions remain over a suitable finite-$N$ expression for the mean self-overlap and various other results at strong and weak non-Hermiticity.

Additionally, it is natural to wish to build upon results for the mean self-overlap by considering higher moments of $\mathcal{O}_{nn}$, or even the entire distribution. For example, the finite-$N$ distribution of self-overlaps associated with complex eigenvalues remains an open problem in the: GinOE, GinSE and all elliptic Ginibre ensembles. Finally, one can also consider statistics of the off-diagonal entries of the overlap matrix, Eq. \eqref{eq:O_def}, so as to ascertain if there is a universality that can be linked to pre-existing results found, for example, in \cite{CM,BD,AFK}.

\section{Derivation of Main Results}
\label{sec:proofs_eGinOUE_edge}

\subsection{Preliminaries for Analysis}

Before we go on to prove our main results, it is convenient for us to briefly introduce some important identities that will be commonly utilised throughout our proofs. The first of these is the following integral representations of the Hermite polynomials,
\begin{equation}
\begin{aligned}
    &\He_k\left( \frac{\bar{z}}{\sqrt{\tau}} \right) = \frac{i^k}{\sqrt{2 \pi \tau}} \exp{\frac{\bar{z}^2}{2 \tau}} \int_{-\infty}^\infty dt_1 \exp{ -\frac{t_1^2}{2 \tau} - i \frac{\bar{z}}{\tau} t_1 } \,  t_1^k \,  \tau^{-k/2}
    \label{eq:He_k_int}
\end{aligned}
\end{equation}
and the corresponding representation of $\HE_k(z/\sqrt{\tau})$, where we choose to exchange $i \to - i$. Upon introducing these integral representations in all proofs, one must then utilise the incomplete $\Gamma$-function, defined as
\begin{equation}
    \Gamma\left( N + 1, x \right) = \Gamma\left(N+1\right) \, e^{-x} \, \sum_{k=0}^{N} \frac{x^k}{k!} =  x^{N+1}\int_1^\infty du \, u^N e^{-ux} \ ,
    \label{eq:incmpl_Gamma}
\end{equation}
where $\Gamma(N) = (N-1)!$. Finally, for our proofs at weak non-Hermiticity, it is useful to employ the following asymptotic behaviour of the incomplete $\Gamma$-function 
\begin{equation}
    \Gamma(m+1,x) = e^{-x} \frac{x^{m+1}}{x-m} \left[ 1 + O\left( \frac{m}{(x-m)^2} \right) ... \right] \ , 
    \label{eq:asym_incmpl_Gamma}
\end{equation}
valid for $x \gtrsim m$ and $m \gg 1$, cf. \cite{GGNV}.

\subsection{Strong non-Hermiticity proofs}

\subsubsection{Proof of Theorem \ref{thm:O_eGinOUE} - Self-Overlap in the eGinUE}

Over the course of the next two sections, we provide a proof of the universality of the leading order behaviour of the mean self-overlap of eigenvectors at strong non-Hermiticity in the edge regime of the eGinUE and eGinOE. Focusing first on the eGinUE, we begin by analysing the finite-$N$ equation for the mean self-overlap, given in Eq. \eqref{eq:O_N_eGinUE}. Since $1 - \tau^2 \sim O(1)$ and $\rho_N^{\textup{(eGinUE,c)}}(z) \sim O(1)$ as $N \to \infty$ at SNH, cf. \cite{LR}, we see that the mean self-overlap in this region can be obtained through
\begin{equation}
    \lim_{N \to \infty} \mathcal{O}_N^{\textup{(eGinUE,c)}}(z) = (1 - \tau^2) \lim_{N \to \infty} \left[ N \rho_{N+1}^{\textup{(eGinUE,c)}}(z) - R_N \right] \ .
\end{equation}
At this point, we introduce the integral representations of the Hermite polynomials from Eq. \eqref{eq:He_k_int} and evaluate the summations using the definition of the incomplete $\Gamma$-function in Eq. \eqref{eq:incmpl_Gamma}, thus
\begin{equation}
\begin{aligned}
    \mathcal{O}_N^{\textup{(eGinUE,c)}}(z) \approx & \frac{\sqrt{1 - \tau^2}}{2 \pi^2 \tau \Gamma(N)} \exp{- \frac{|z|^2 - \tau \Re(z^2)}{1 - \tau^2} + \frac{z^2 + \bar{z}^2}{2 \tau}} \int_{-\infty}^\infty dt_1 \int_{-\infty}^\infty dt_2 \\
    &\exp{ t_1 t_2 - \frac{t_1^2 + t_2^2}{2 \tau} + \frac{i}{\tau} \left( z t_2 - \bar{z} t_1 \right)  } \Big( \Gamma(N+1, t_1 t_2) - t_1 t_2 \Gamma(N, t_1 t_2) \Big) \ .  
\end{aligned}
\end{equation}
One may now make use of the integral representation of the incomplete $\Gamma$-function in Eq. \eqref{eq:incmpl_Gamma}, which leads to 
\begin{equation}
\begin{aligned}
    \mathcal{O}_N^{\textup{(eGinUE,c)}}(z) \approx & \frac{\sqrt{1 - \tau^2}}{2 \pi^2 \tau \Gamma(N)} \exp{- \frac{|z|^2 - \tau \Re(z^2)}{1 - \tau^2} + \frac{z^2 + \bar{z}^2}{2 \tau}} \int_1^\infty du \, \frac{u-1}{u} \int_{-\infty}^\infty dt_1 \int_{-\infty}^\infty dt_2 \, \,  t_1 t_2 \\
    &(u t_1 t_2)^{N} \exp{ (1 - u) t_1 t_2 - \frac{t_1^2 + t_2^2}{2 \tau} + \frac{i}{\tau} \left( z t_2 - \bar{z} t_1 \right)  }  \ .  
\end{aligned}
\end{equation}
We now make the substitution $\sqrt{N} p = t_1 + t_2$ and $\sqrt{N} q = t_1 - t_2$, such that
\begin{equation}
    t_1^2 + t_2^2 = \frac{N(p^2 + q^2)}{2} \ , \hspace{1cm}  t_1 \,  t_2 = \frac{N(p^2 - q^2)}{4} \ , \hspace{1cm} dt_1 \, dt_2 = \frac{N}{2} \, dp \, dq \ ,
    \label{eq:p_q_variables}
\end{equation}
which yields
\begin{align}
    \mathcal{O}&_N^{\textup{(eGinUE,c)}}(z) \approx \frac{N^{N+2}\sqrt{1 - \tau^2}}{4 \pi^2 \tau \Gamma(N)} e^{- \frac{|z|^2 - \tau \Re(z^2)}{1 - \tau^2} + \frac{z^2 + \bar{z}^2}{2 \tau}} \int_1^\infty du \, \frac{u-1}{u} \int_{-\infty}^\infty dp \int_{-\infty}^\infty dq \, \frac{p^2 - q^2}{4} e^{ - \frac{i \sqrt{N}w_x q + \sqrt{N} w_y p}{\tau}  } \nonumber  \\
    & \exp{ N \left[ \frac{(1 - u)(p^2 - q^2)}{4} - \frac{p^2 + q^2}{4 \tau} - \frac{i(1+\tau) \cos(\theta)q}{\tau} - \frac{(1-\tau) \sin(\theta) p}{\tau} + \ln(\frac{u(p^2 - q^2)}{4}) \right]} \ , 
    \label{eq:O_edge_eGinUE_pre_exp}
\end{align}
where $w_x$ and $w_y$ are defined in Eq. \eqref{eq:edge_xy_params}. To obtain leading order asymptotic behaviour of the above equation, one must locate the maximum of the function in square brackets. We find that the dominant extremal point is given by $\left( p^*, q^*, u^* \right) = \left(-2 \sin(\theta), -2i \cos(\theta), 1 \right)$, thus, by slightly deforming the contour, we expand around the maximum using
\begin{equation}
    p = - 2 \sin(\theta) + \frac{r}{\sqrt{N}} \hspace{1cm} q = - 2 i \cos(\theta) + \frac{s}{\sqrt{N}} \hspace{1cm} u = 1 + \frac{v}{\sqrt{N}} \ . 
    \label{eq:pqu_subs}
\end{equation}
The expansion around this maximum within the leading exponential is a rather cumbersome process and so we relegate the details to Appendix \ref{app:expo_SNH}. Upon completion of this procedure, we utilise Eq. \eqref{eq:app_final_eq} and the fact that $(p^2 - q^2)/4 \to 1 + O(N^{-1/2})$, to find that 
\begin{align}
    \mathcal{O}_N^{\textup{(eGinUE,c)}}(z) \approx& \frac{\sqrt{1 - \tau^2}}{4 \pi^2 \tau} \frac{N^N e^{-N}}{\Gamma(N)} e^{ - \frac{|z|^2 - \tau \Re(z^2)}{1 - \tau^2} + \frac{z^2 + \bar{z}^2}{2 \tau} + \frac{2\sqrt{N}(w_y \sin(\theta) - w_x \cos(\theta))}{\tau} - \frac{N(\tau + \cos(2\theta))}{\tau}} \nonumber \\
    &\int_0^\infty dv \, v \, e^{-\frac{v^2}{2}} \int_{-\infty}^\infty d r \int_{-\infty}^\infty ds \exp{  - \frac{r^2( 1 - \tau \cos(2\theta))}{4 \tau} - \frac{w_y r}{\tau} + rv \sin(\theta)} \label{eq:O_eGinUE_SNH_coupled} \\ 
    &\exp{ - \frac{s^2( 1 - \tau \cos(2\theta))}{4 \tau} - \frac{i w_x s}{\tau} - i s v \cos(\theta) + i r s \cos(\theta) \sin(\theta) } \nonumber \ .
\end{align}
Therefore, one must calculate the coupled Gaussian integrals over $r$ and $s$ using the following identity
\begin{equation}
    \int_{-\infty}^\infty dr \, \exp{-\alpha r^2 + \beta r} \int_{-\infty}^\infty ds \, \exp{-a s^2 + b s + crs} = \frac{2\pi}{\sqrt{4 \alpha a - c^2}} \exp{\frac{\alpha b^2 + b \beta c + a \beta ^2}{4 \alpha a - c^2}} \ ,
    \label{eq:CGI_start}
\end{equation}
where in our case, with the help of Mathematica, the prefactor and exponent can be evaluated as
\begin{equation}
    \frac{2\pi}{\sqrt{4 \alpha a - c^2}} = \frac{4 \pi \tau}{ \sqrt{1 + \tau^2 - 2 \tau \cos(2 \theta)} } 
\end{equation}
and
\begin{equation}
    \frac{\alpha b^2 + b \beta c + a \beta ^2}{4 \alpha a - c^2} = \frac{\eta^2\Big( \tau - \cos(2 \theta) \Big) - 2 v \eta \tau \sqrt{ 1 + \tau^2 - 2 \tau \cos(2 \theta)} + v^2 \tau^2 \Big(\tau - \cos(2 \theta)\Big) }{\tau (1 + \tau^2 - 2\tau \cos(2\theta))} \ ,
    \label{eq:CGI_end}
\end{equation}
respectively. Stirling's formula ($\Gamma(N) = \sqrt{2\pi}N^{N-1/2}e^{-N}$) can also be applied and one can simplify the exponential prefactor using Mathematica. Eventually this produces
\begin{equation}
\begin{aligned}
    \mathcal{O}_N^{\textup{(eGinUE,c)}}(z) \approx& \frac{1}{\pi} \sqrt{\frac{N}{2 \pi}\frac{1 - \tau^2}{1 + \tau^2 - 2 \tau \cos(2 \theta)}} e^{- \frac{\eta^2 ( \tau (\tau^2 + 3) - (1 + 3 \tau^2) \cos(2 \theta) )}{\tau(1 - \tau^2)(1 + \tau^2 - 2 \tau \cos(2 \theta))} + \frac{\eta^2(\tau - \cos(2 \theta))}{\tau(1 + \tau^2 - 2 \tau \cos(2 \theta) )}}\\
    &\int_0^\infty dv \, v \, \exp{ - \frac{v^2}{2} + \frac{v^2 \tau(\tau - \cos(2 \theta))}{(1 + \tau^2 - 2\tau \cos(2\theta))} - \frac{  2 v \eta   }{\sqrt{1 + \tau^2 - 2\tau \cos(2\theta)}}  } \ .
\end{aligned}
\end{equation}
One may now complete the square on the terms inside the exponential of the final integral, such that 
\begin{equation}
\begin{aligned}
    - \frac{v^2}{2}  + \frac{v^2 \tau(\tau - \cos(2 \theta))}{(1 + \tau^2 - 2\tau \cos(2\theta))} -& \frac{  2 v \eta   }{\sqrt{1 + \tau^2 - 2\tau \cos(2\theta)}}\\ 
    &= - \frac{1 - \tau^2}{2(1 + \tau^2 - 2 \tau \cos(2 \theta))} \left( v+ \frac{2 \eta \sqrt{1 + \tau^2 - 2 \tau \cos(2 \theta) }}{1 - \tau^2} \right)^2 + \frac{2 \eta^2}{1 - \tau^2} \label{eq:comp_square}
\end{aligned}   
\end{equation}
and upon employing the identity
\begin{equation}
    \int_0^\infty dx \, x \exp{-\frac{a}{2}(x+b)^2} = \frac{1}{a} \exp{- \frac{ab^2}{2}}  - b\sqrt{\frac{\pi}{2a}} \erfc\left(  b \sqrt{\frac{a}{2}}\right) \ ,
    \label{eq:final_int_O_edge}
\end{equation}
it becomes clear that
\begin{align}
    \mathcal{O}_N^{\textup{(eGinUE,c)}}(z) \approx& \frac{1}{\pi} \sqrt{\frac{N}{2\pi} \frac{1 - \tau^2}{1 + \tau^2 - 2 \tau \cos(2 \theta)} } \int_0^\infty dv \, v \, e^{- \frac{1 - \tau^2}{2(1 + \tau^2 - 2 \tau \cos(2 \theta))} \left( v+ \frac{2 \eta \sqrt{1 + \tau^2 - 2 \tau \cos(2 \theta) }}{1 - \tau^2} \right)^2 }\\
    =& \frac{\sqrt{N}}{\pi} \sqrt{\frac{1 + \tau^2 - 2 \tau \cos(2 \theta)}{1 - \tau^2}} \left[ \frac{1}{\sqrt{2 \pi}} \exp{  \frac{-2\eta^2}{1 - \tau^2}} - \frac{\eta}{\sqrt{{1 - \tau^2}}} \textup{erfc}\left( \sqrt{\frac{2}{1 - \tau^2}} \eta \right)\right] \ .
\end{align}
Thus, by a simple rescaling and taking the limit of $N \to \infty$ we arrive at Eq. \eqref{eq:O_SNH_edge_eGinUE}, concluding the proof.

\subsubsection{Proof of Theorem \ref{thm:O_eGinOUE} - Self-Overlap in the eGinOE}

The proof of Theorem \ref{thm:O_eGinOUE} in the eGinOE follows steps that are similar to the ones outlined in the previous two proofs. With that in mind, we proceed to highlight some key points of the calculation, dealing with large amounts of the technical details through the use of previous material. We begin by considering the form of the finite-$N$ result for the mean self-overlap presented in Eq. \eqref{eq:O_N_eGinOE}. Immediately, one can note that, in the limit of large-$N$, the term containing the error function becomes negligible and the dominant term within the square bracket is $N P_N - T_N$, therefore,
\begin{equation}
    \lim_{N \to \infty} \mathcal{O}_N^{\text{(eGinOE,c)}}(z) = \frac{(1 - \tau) \sqrt{1 - \tau^2}}{\pi} \lim_{N \to \infty} \left[ \exp{- \frac{|z|^2 - \tau \Re(z^2)}{1 - \tau^2}} \Big( N P_N - T_N \Big) \right]\ .
\end{equation}
Using the definitions of $P_N$ and $T_N$, Eqs. \eqref{eq:P_N} and \eqref{eq:T_N} respectively, the integral representations of Hermite polynomials, Eq. \eqref{eq:He_k_int}, and the series representation of the incomplete $\Gamma$-function, Eq. \eqref{eq:incmpl_Gamma}, one finds that
\begin{equation}
\begin{aligned}
    \mathcal{O}_N^{\text{(eGinOE,c)}}(z) \approx& \frac{(1 - \tau) \sqrt{1 - \tau^2}}{4\pi^2 \tau \, \Gamma(N)} \left( \frac{-1}{\IM(z)} \right) \exp{- \frac{|z|^2 - \tau \Re(z^2)}{1 - \tau^2} + \frac{z^2 + \bar{z}^2}{2 \tau}}\int_{-\infty}^\infty dt_1 \int_{-\infty}^\infty dt_2 \\
    &(t_1 + t_2) \, \exp{ t_1 t_2 - \frac{t_1^2 + t_2^2}{2 \tau} + \frac{i}{\tau} \left( z t_2 - \bar{z} t_1 \right)  } \Big( \Gamma(N+1, t_1 t_2) - t_1 t_2 \Gamma(N, t_1 t_2) \Big) \ .
\end{aligned}
\end{equation}
Introducing integral representations of the incomplete $\Gamma$-functions, according to Eq. \eqref{eq:incmpl_Gamma}, one then arrives at
\begin{equation}
\begin{aligned}
    \mathcal{O}_N^{\text{(eGinOE,c)}}(z) \approx& \frac{(1 - \tau) \sqrt{1 - \tau^2}}{4\pi^2 \tau \, \Gamma(N)} \left( \frac{-1}{\IM(z)} \right) \exp{- \frac{|z|^2 - \tau \Re(z^2)}{1 - \tau^2} + \frac{z^2 + \bar{z}^2}{2 \tau}} \int_0^\infty du \, \frac{u-1}{u} \\
    & \int_{-\infty}^\infty dt_1 \int_{-\infty}^\infty dt_2 \, (t_1 + t_2) \, t_1 t_2 \, (u t_1 t_2)^N \, \exp{ t_1 t_2 - \frac{t_1^2 + t_2^2}{2 \tau} + \frac{i}{\tau} \left( z t_2 - \bar{z} t_1 \right)  }  
\end{aligned}
\end{equation}
and upon employing the usual substitutions of $\sqrt{N} p = t_1 + t_2$ and $\sqrt{N} q = t_1 - t_2$, we find that
\begingroup
\allowdisplaybreaks
\begin{align}
    \mathcal{O}&_N^{\text{(eGinOE,c)}}(z) \approx \frac{-N^{N+5/2}(1 - \tau) \sqrt{1 - \tau^2}}{8\pi^2 \tau \, \Gamma(N) \, \IM(z) }  e^{- \frac{|z|^2 - \tau \Re(z^2)}{1 - \tau^2} + \frac{z^2 + \bar{z}^2}{2 \tau}} \int_0^\infty du \, \frac{u-1}{u} \int_{-\infty}^\infty dp \int_{-\infty}^\infty dq \, \frac{p(p^2 - q^2)}{4} \label{eq:O_edge_eGinOE_pre_exp}\\
    &e^{ - \frac{i \sqrt{N}w_x q}{\tau} - \frac{\sqrt{N} w_y p}{\tau} } \exp{ N \left[ \frac{(1 - u)(p^2 - q^2)}{4} - \frac{p^2 + q^2}{4 \tau} - \frac{i(1+\tau) \cos(\theta)q}{\tau} - \frac{(1-\tau) \sin(\theta) p}{\tau} + \ln(\frac{u(p^2 - q^2)}{4}) \right]} \ , \nonumber
\end{align}
\endgroup
with $w_x$ and $w_y$ defined in Eq. \eqref{eq:edge_xy_params}. As in the previous sections, this integral is dominated at the maximum of the final exponential located at $\left( p^*, q^*, u^* \right) = \left(-2 \sin(\theta), -2i \cos(\theta), 1 \right)$. For this reason, we expand around the maximum using the substitutions defined in Eq. \eqref{eq:pqu_subs}, as detailed in Appendix \ref{app:expo_SNH}, yielding
\begingroup
\allowdisplaybreaks
\begin{align}
    \mathcal{O}_N^{\text{(eGinOE,c)}}(z) \approx& \frac{N^{N+1/2}(1 - \tau) \sqrt{1 - \tau^2}\sin(\theta)}{8\pi^2 \tau \, \Gamma(N) \, \IM(z) }  e^{- \frac{|z|^2 - \tau \Re(z^2)}{1 - \tau^2} + \frac{z^2 + \bar{z}^2}{2 \tau} \frac{2\sqrt{N}(w_y \sin(\theta) - w_x \cos(\theta))}{\tau} - \frac{N(\tau + \cos(2\theta))}{\tau} } \nonumber \\
    &\int_0^\infty dv \, v \, e^{-\frac{v^2}{2}} \int_{-\infty}^\infty d r \int_{-\infty}^\infty ds \exp{  - \frac{r^2( 1 - \tau \cos(2\theta))}{4 \tau} - \frac{w_y r}{\tau} + rv \sin(\theta)}  \\ 
    &\exp{ - \frac{s^2( 1 - \tau \cos(2\theta))}{4 \tau} - \frac{i w_x s}{\tau} - i s v \cos(\theta) + i r s \cos(\theta) \sin(\theta) } \nonumber \ .
\end{align}
\endgroup
To complete this derivation, we evaluate the coupled Gaussian integrals using exactly the same procedure as in Eqs. \eqref{eq:CGI_start} to \eqref{eq:CGI_end} and we simplify the prefactors using Stirling's formula and $\IM(z) \approx \sqrt{N}(1 - \tau) \sin(\theta)$. One can then complete the square on the resulting exponential using Eq. \eqref{eq:comp_square}, yielding
\begingroup
\allowdisplaybreaks
\begin{align}
    \mathcal{O}_N^{\textup{(eGinOE,c)}}(z) \approx& \frac{1}{\pi} \sqrt{\frac{N}{2\pi} \frac{1 - \tau^2}{1 + \tau^2 - 2 \tau \cos(2 \theta)} } \int_0^\infty dv \, v \, e^{- \frac{1 - \tau^2}{2(1 + \tau^2 - 2 \tau \cos(2 \theta))} \left( v+ \frac{2 \eta \sqrt{1 + \tau^2 - 2 \tau \cos(2 \theta) }}{1 - \tau^2} \right)^2 }\\
    =& \frac{\sqrt{N}}{\pi} \sqrt{\frac{1 + \tau^2 - 2 \tau \cos(2 \theta)}{1 - \tau^2}} \left[ \frac{1}{\sqrt{2 \pi}} \exp{\frac{-2\eta^2}{1 - \tau^2} } - \frac{\eta}{\sqrt{{1 - \tau^2}}} \textup{erfc}\left( \sqrt{\frac{2}{1 - \tau^2}} \eta \right)\right] \ ,
\end{align}
\endgroup
where the final integral has been evaluated using Eq. \eqref{eq:final_int_O_edge}. Thus, by rescaling and taking the limit of $N \to \infty$ we arrive at the desired result of Eq. \eqref{eq:O_SNH_edge_eGinOE}.

\subsection{Weak non-Hermiticity proofs}

\subsubsection{Proof of Proposition \ref{prop:rho_eGinOE_WNH_edge} - Density in the eGinOE}
\label{sec:rho_eGinOE_WNH_edge}

We now move on to prove the results pertaining to the weak non-Hermiticity regime, outlined in Section \ref{sec:WNH_edge_results}, and begin by considering the density of complex eigenvalues at the edge for the eGinOE. In order to derive the result in Eq. \eqref{eq:rho_eGinOE_WNH_edge}, one starts from the finite-$N$ equation in Eq. \eqref{eq:rho_N_eGinUE} and can immediately say that 
\begin{equation}
\begin{aligned}
    \rho_N^{\text{(eGinOE,c)}}(z) \approx & \frac{i}{2\sqrt{2\pi}} \ \erfc\left( \frac{y}{\sqrt{\kappa}} \right) \, \exp{ \frac{\IM(z)^2 - \RE(z)^2}{1 + \tau} } \\
    & \sum_{k=0}^N \frac{\tau^{k+\frac{1}{2}}}{k!} \left[ \HE_{k+1}\left( \frac{\bar{z}}{\sqrt{\tau}} \right)  \HE_{k}\left( \frac{z}{\sqrt{\tau}} \right) - \HE_{k+1}\left( \frac{z}{\sqrt{\tau}} \right) \HE_{k}\left( \frac{\bar{z}}{\sqrt{\tau}} \right)\right] \ ,
    \label{eq:WNH_steps_start}
\end{aligned}
\end{equation}
where we have utilised the definition of $P_N$ and the parameterisation of the edge at WNH from Eqs. \eqref{eq:P_N} and \eqref{eq:edge_xy_params} respectively. One can now introduce the integral representations of the Hermite polynomials from Eq. \eqref{eq:He_k_int} and carry out the summation using the definition of the incomplete $\Gamma$-function from Eq. \eqref{eq:incmpl_Gamma}, yielding
\begin{equation}
\begin{aligned}
    \rho_N^{\text{(eGinOE,c)}}(z) \approx&  \frac{-1}{2(2 \pi)^{3/2}} \erfc\left( \frac{y}{\sqrt{\kappa}} \right) \exp{ \frac{\IM(z)^2 - \RE(z)^2}{1 + \tau} + \frac{z^2 + \bar{z}^2}{ 2 \tau } }  \\
    & \int_{-\infty}^\infty dt_1 \int_{-\infty}^\infty dt_2 \, (t_1 + t_2) \, \exp{t_1 t_2 - \frac{t_1^2 + t_2^2}{2 \tau} - \frac{i \bar{z}t_1}{\tau} + \frac{iz t_2}{\tau} } \,  \frac{\Gamma(N+1, t_1 t_2)}{\Gamma(N+1)} \ .
\end{aligned}
\end{equation}
In order to make the above equation more amenable to saddle point analysis, we rescale $(t_1,t_2) \to \sqrt{N}(t_1, t_2)$ and employ the asymptotic behaviour of the incomplete $\Gamma$-function, given in Eq. \eqref{eq:asym_incmpl_Gamma}, producing
\begin{equation}
\begin{aligned}
   \rho_N^{\text{(eGinOE,c)}}(z) \approx& \left( \frac{-N^{N+1/2} }{2 (2 \pi)^{3/2} \Gamma(N)} \right) \erfc\left( \frac{y}{\sqrt{\kappa}} \right) \exp{ \frac{\IM(z)^2 - \RE(z)^2}{1 + \tau} + \frac{z^2 + \bar{z}^2}{ 2 \tau } } \int_{-\infty}^\infty dt_1 \int_{-\infty}^\infty dt_2 \\
    & \frac{(t_1 + t_2) t_1 t_2}{t_1 t_2 - 1} \exp{- N \left( \frac{t_1^2}{2 \tau} + \frac{i \bar{z} t_1}{\sqrt{N}\tau} - \ln(t_1) \right) } \exp{-N \left( \frac{t_2^2}{2 \tau} - \frac{i z t_2}{\sqrt{N}\tau} - \ln(t_2) \right) } \ .
    \label{eq:WNH_steps_end}
\end{aligned}
\end{equation} 
Therefore, one must locate and expand around the saddle points of
\begin{equation}
    g_1(t_1) = \frac{t_1^2}{2\tau} + \frac{i \bar{z} t_1}{\tau \sqrt{N}} - \ln(t_1) \hspace{1cm} \text{and} \hspace{1cm} g_2(t_2) = \frac{t_2^2}{2\tau} - \frac{i z t_2}{\tau \sqrt{N}} - \ln(t_2) \ . 
\end{equation}
Analysis of the above functions yields that the saddle point is located at $(t_1^*, t_2^*) = (-i,i)$ and that the second derivatives both vanish there, i.e. $g_1''(t_1^*) = g_2''(t_2^*) = 0$. Hence, we must expand around the maximum using
\begin{equation}
    t_1 = - i - \frac{p}{N^{1/3}} \hspace{2cm} \text{and} \hspace{2cm}  t_2 =  i + \frac{q} {N^{1/3}} \ .
    \label{eq:t1_t2_p_q_sub}
\end{equation}
With the help of Mathematica, it is possible to carefully Taylor expand the terms inside the exponential and, after some algebraic steps, we find that 
\begin{equation}
\begin{aligned}
     \frac{\IM(z)^2 - \RE(z)^2}{1 + \tau} +& \frac{z^2 + \bar{z}^2}{ 2 \tau } - N \left( \frac{t_1^2}{2 \tau} + \frac{i \bar{z} t_1}{\sqrt{N}\tau} - \ln(t_1) \right) -N \left( \frac{t_2^2}{2 \tau} - \frac{i z t_2}{\sqrt{N}\tau} - \ln(t_2) \right) \\
     & \to  - N + \frac{ip^3}{3} +ipx + \frac{iq^3}{3} +iqx + \frac{\frac{x^2}{2} + (p-q) y  + i (p+q) \kappa }{ N^{1/3} } + O(N^{-2/3}) \ .
\end{aligned}
\end{equation}
Additionally, it is easy to see that the remaining part of the integrand can be written as
\begin{equation}
    \frac{(t_1 + t_2) t_1 t_2}{t_1 t_2 - 1} \to \frac{p-q}{i(p+q)} + O(N^{-1/3}) \ .
    \label{eq:pre_expo_t1t2_pq}
\end{equation}
Returning to the density, this means that
\begin{equation}
\begin{aligned}
    \rho_N^{\text{(eGinOE,c)}}(z) \approx& \left( \frac{-N^{N-1/6} e^{-N} }{2 (2 \pi)^{3/2} \Gamma(N) } \right) \erfc\left( \frac{y}{\sqrt{\kappa}} \right) \int_{-\infty}^\infty dp \int_{-\infty}^\infty dq \, \frac{p-q}{i(p+q)} \\
    &\exp{ \frac{ip^3}{3} + ipx + \frac{iq^3}{3} + iqx} \left( 1+ \frac{\frac{x^2}{2} + (p-q) y  + i (p+q) \kappa }{ N^{1/3} } \right) 
\end{aligned}
\end{equation}
and one can see that the only term which contributes to this integral is the term proportional to $(p-q)^2$. Therefore, we obtain
\begin{equation}
\begin{aligned}
    \rho_N^{\text{(eGinOE,c)}}(z) \approx& \frac{-1}{2 (2 \pi)^{2}} y \, \erfc\left( \frac{y}{\sqrt{\kappa}} \right) \int_{-\infty}^\infty dp \int_{-\infty}^\infty dq \, \frac{(p-q)^2}{i(p+q)} \exp{ \frac{ip^3}{3} + ipx + \frac{iq^3}{3} + iqx}  \ , 
\end{aligned}
\end{equation}
where Stirling's formula has been utilised to simplify the prefactors. If one now makes the change of variables $(p,q) \to i(p,q)$ and deforms the contour from along the imaginary axis to a curved contour, $\cont$, starting at infinity with an argument of $-\pi/3$ and ending at infinity with an argument of $\pi/3$, it can be seen that
\begin{equation}
\begin{aligned}
    \rho_N^{\text{(eGinOE,c)}}(z) \approx& \frac{1}{2 (2 \pi)^{2}} y \, \erfc\left( \frac{y}{\sqrt{\kappa}} \right) \int_{\cont} dp \int_{\cont} dq \, \frac{(p-q)^2}{p+q} \exp{  \frac{p^3}{3} - px + \frac{q^3}{3} - qx }  \ . 
\end{aligned}
\end{equation}
Utilising $a^{-1} = \int_0^\infty dt \, e^{-at}$, for $\Re(a)>0$, it becomes clear that
\begin{align}
    \rho_N^{\text{(eGinOE,c)}}(z) \approx&  \frac{-1}{2}  y\, \erfc\left( \frac{y}{\sqrt{\kappa}} \right) \frac{1}{(2\pi i)^2} \int_0^\infty dt \int_{\cont} dp \int_{\cont} dq \, (p-q)^2 e^{ \frac{p^3}{3} - p(x+t) + \frac{q^3}{3} - q(x+t)} \\
    =&  \, y \, \erfc\left( \frac{y}{\sqrt{\kappa}} \right) \int_0^\infty dt \bigg[ \Ai'(x+t)^2 - \Ai(x+t) \, \Ai''(x+t) \bigg] \ ,
\end{align}
where $\Ai'(\cdot)$ and $\Ai''(\cdot)$ are the first and second derivatives of the Airy function, defined in Eq. \eqref{eq:Airy_func}, with general $k$-th derivative
\begin{equation}
    \Ai^{(k)}(z) = \frac{\partial^k}{\partial x^k} \Ai(x) \bigg|_{x = z} = \frac{(-1)^k}{2\pi i} \int_{\cont} d p \, p^{k} \exp{ \frac{p^3}{3} - pz }  \ .
    \label{eq:Ai_derivatives}
\end{equation}

\subsubsection{Proof of Theorem \ref{thm:S_eGinUE_WNH_edge} - Self-Overlap in the eGinUE}

We now move on to the mean self-overlap of eigenvectors at the edge for the eGinUE at weak non-Hermiticity. Considering the finite-$N$ form of the mean self-overlap, Eq. \eqref{eq:O_N_eGinUE}, one can see that the term proportional to $\rho_{N-1}^{\textup{(eGinUE,c)}}(z)$ cannot possibly contribute to the leading order asymptotic behaviour, since $1 - \tau^2 \sim O(1/N)$. Following a very similar method to the previous section, one can also see that, to leading order, $R_N \approx N \rho_{N+1}$. Therefore, at the edge of the eGinUE in the weak non-Hermiticity regime, $\mathcal{O} \approx \rho$ and consequently $\mathbb{E} \to 1$ as $N \to \infty$. For this reason, we choose to consider the leading order deviation from $\mathbb{E} = 1$, quantified by
\begin{equation}
    \delta_N = (1 - \tau^2) \Big( N \rho_{N+1}^{\textup{(eGinUE,c)}}(z) - R_N \Big) \ ,
\end{equation}
such that for large-$N$, $\mathcal{O}_N \approx \rho_N + \delta_N$. Using the definitions of $\rho_{N}^{\textup{(eGinUE,c)}}(z) $ and $ R_N$ from Eqs. \eqref{eq:rho_N_eGinUE} and \eqref{eq:R_N}, introducing the integral representations of the Hermite polynomials from Eq. \eqref{eq:He_k_int} and evaluating the summation using the definition of the incomplete $\Gamma$-function in Eq. \eqref{eq:incmpl_Gamma}, one finds that 
\begin{equation}
\begin{aligned}
    \delta_N \approx& \frac{1}{\sqrt{N} \pi^2 } \sqrt{\frac{\kappa}{2}} \frac{1}{\Gamma(N)} \exp{ - \frac{|z|^2 - \tau \RE(z^2)}{1- \tau^2} + \frac{z^2 + \bar{z}^2}{2 \tau}} \int_{-\infty}^\infty dt_1 \int_{-\infty}^\infty dt_2 \\
    & \hspace{1cm} \hspace{1cm} \exp{ t_1 t_2 - \frac{t_1^2 + t_2^2}{2 \tau} - \frac{i \bar{z} t_1}{\tau} + \frac{i z t_2}{\tau}  } \, \bigg[ \Gamma(N+1, t_1 t_2) - t_1 t_2 \, \Gamma(N, t_1 t_2) \bigg] \ .
\end{aligned}
\end{equation}
We now rescale $(t_1, t_2) \to \sqrt{N}(t_1, t_2)$ and then employ the asymptotic behaviour of the incomplete $\Gamma$-function from Eq. \eqref{eq:asym_incmpl_Gamma}, to see that  
\begin{equation}
\begin{aligned}
    \delta_N \approx& \frac{N^{N+3/2}}{\pi^2} \sqrt{\frac{\kappa}{2}} \frac{1}{\Gamma(N)} e^{ - \frac{|z|^2 - \tau \RE(z^2)}{1- \tau^2} + \frac{z^2 + \bar{z}^2}{2 \tau}} \int_{-\infty}^\infty dt_1 \int_{-\infty}^\infty dt_2 \, t_1 t_2 \bigg[ \frac{1}{Nt_1 t_2 - N} - \frac{1}{Nt_1 t_2 - N + 1} \bigg]  \\
    & \exp{- N \left( \frac{t_1^2}{2 \tau} + \frac{i \bar{z} t_1}{\sqrt{N}\tau} - \ln(t_1) \right) } \exp{-N \left( \frac{t_2^2}{2 \tau} - \frac{i z t_2}{\sqrt{N}\tau} - \ln(t_2) \right) } \ .
\end{aligned}
\end{equation}
As previously, these integrals are dominated in the region of $(t_1, t_2) = (-i,i)$, with the second derivative of the functions inside the exponential being zero at the saddle point. Hence, we employ the substitution
\begin{equation}
    t_1 = - i - \frac{p}{N^{1/3}} \hspace{2cm} \text{and} \hspace{2cm}  t_2 =  i + \frac{q} {N^{1/3}} \ ,
\end{equation}
and find that
\begin{equation}
    \delta_N \approx \frac{N^{N - \frac{1}{2}}e^{-N}}{\pi^2 \Gamma(N) } \sqrt{\frac{\kappa}{2}} \exp{\frac{-y^2}{\kappa}}  \int_{-\infty}^\infty dp  \int_{-\infty}^\infty dq \, \frac{e^{\frac{ip^3}{3} + ipx + \frac{iq^3}{3} + i q x}}{(p+q)^2} \ ,
\end{equation}
since
\begin{equation}
\begin{aligned}
    - \frac{|z|^2 - \tau \RE(z^2)}{1 - \tau^2} + & \frac{z^2 + \bar{z}^2}{ 2 \tau } - N \left( \frac{t_1^2}{2 \tau} + \frac{i \bar{z} t_1}{\sqrt{N}\tau} - \ln(t_1) \right) -N \left( \frac{t_2^2}{2 \tau} - \frac{i z t_2}{\sqrt{N}\tau} - \ln(t_2) \right) \\ 
    & \to - N - \frac{y^2}{\kappa} + \frac{ip^3}{3} +ipx + \frac{iq^3}{3} +iqx + O(N^{-1/3})
\end{aligned}
\end{equation}
and
\begin{equation}
    \frac{1}{Nt_1 t_2 - N} - \frac{1}{Nt_1 t_2 - N + 1} = - \frac{1}{(p+q)^2 N^{4/3}} + O(N^{-5/3})\ .  
\end{equation}
Note that one can employ finer asymptotics of the incomplete $\Gamma$-function to verify that higher order terms in the expansion do not contribute to the leading term of the above equation, see e.g. \cite[Eq. (8.11.7)]{NIST}. One now rescales $(p,q) \to i (p,q)$, deforms the contours of the $p$ and $q$ integrals to ${\cont}$ and utilises $a^{-2} = \int_0^\infty dt \, t \, e^{-at}$, for $\Re(a) > 0$. This leads to 
\begin{equation}
    \delta_N \approx 2\sqrt{\frac{ \kappa}{\pi}} \exp{ \frac{-y^2}{\kappa} }  \int_0^\infty dt \, t \, \Ai(x+t)^2
\end{equation}
and so by introducing $\mathcal{S}_N = (\mathcal{O}_N - \rho_N)/\rho_N = \mathbb{E}_N - 1 \approx \delta_N / \rho_N$, we find that
\begin{equation}
\begin{aligned}
    \mathcal{S}^{\textup{(eGinUE,c)}}_{\textup{WNH, edge}}(x) =& \lim_{N \to \infty} N^{2/3} \bigg( \mathbb{E}^{\textup{(eGinUE,c)}}_{N} \left( z = 2 \sqrt{N} + \frac{x}{N^{1/6}} + i \frac{y}{N^{1/2}} \right) - 1 \bigg) = \frac{2 \kappa \int_0^\infty dt \, t \, \textup{Ai}(x+t)^2}{\int_0^\infty dt  \, \textup{Ai}(x+t)^2} \ .
\end{aligned}
\end{equation}

\subsubsection{Proof of Theorem \ref{thm:E_eGinOE_WNH_edge} - Self-Overlap in the eGinOE}

The proof of our final Theorem relies heavily on the techniques outlined in Section \ref{sec:rho_eGinOE_WNH_edge}. For this reason we only outline key differences in this section and leave large amounts of the technical detail as an exercise for the reader. Considering the finite-$N$ form of the mean self-overlap, Eq. \eqref{eq:O_N_eGinOE}, it can be seen that, at the edge, as $N$ becomes large, $\mathcal{O}_N$ does not approach $\rho_N$ in the eGinOE. This is in contrast to what we observed in the eGinUE. Initially, one utilises Eq. \eqref{eq:z_edge_WNH} in the prefactors to note that
\begin{equation}
    \Bigg[ 1 + \sqrt{\frac{\pi(1-\tau^2)}{2}} \ \exp{  \frac{2 \IM(z)^2}{1-\tau^2} } \, \frac{1}{2\vert \IM(z) \vert} \ \textup{erfc}\left( \sqrt{\frac{2 }{1-\tau^2}} \ \vert \IM(z) \vert \right) \Bigg] \to  1 + \frac{\sqrt{\pi \kappa}}{2 y} \erfc\left( \frac{y}{\sqrt{\kappa}} \right) \ .
\end{equation}
Then, following a process similar to Section \ref{sec:rho_eGinOE_WNH_edge}, we find that, to leading order, $N P_N \sim T_N$, therefore the leading order behaviour of the mean self-overlap is given by
\begin{equation}
    \mathcal{O}_N^{\text{(eGinOE,c)}}(z) \approx \left[ 1 + \frac{\sqrt{\pi \kappa}}{2 y} \exp{  \frac{y^2}{\kappa} } \erfc\left( \frac{y}{\sqrt{\kappa}} \right) \right] Q_N \ ,
\end{equation}
where we have introduced
\begin{equation}
    Q_N = \frac{1}{\pi} \ \sqrt{\frac{1-\tau}{1+\tau}} \exp{-\frac{\RE(z)^2}{1+\tau} -\frac{\IM(z)^2}{1-\tau} } P_N \ .
\end{equation}
At this point, we introduce the definition of $P_N$ and the integral representations of the Hermite polynomials, then carry out the summation, in exactly the same way as in Eqs. \eqref{eq:WNH_steps_start} to \eqref{eq:WNH_steps_end}, eventually yielding
\begin{equation}
\begin{aligned}
    Q_N \approx& \frac{N^{N+1/2}}{(2 \pi i)^2} \sqrt{\frac{\kappa}{2}} \frac{1}{y \Gamma(N)} \exp{-\frac{\RE(z)^2}{1+\tau} - \frac{\IM(z)^2}{1-\tau} + \frac{z^2 + \bar{z}^2}{2 \tau}} \int_{-\infty}^\infty dt_1 \int_{-\infty}^\infty dt_2 \\
    & \frac{(t_1 + t_2) t_1 t_2}{t_1 t_2 - 1} \exp{- N \left( \frac{t_1^2}{2 \tau} + \frac{i \bar{z} t_1}{\sqrt{N}\tau} - \ln(t_1) \right) } \exp{-N \left( \frac{t_2^2}{2 \tau} - \frac{i z t_2}{\sqrt{N}\tau} - \ln(t_2) \right) } \ .
\end{aligned}
\end{equation}
Once again, this double integral is dominated in the vicinity of $(t_1,t_2) = (-i,i)$ and so we employ the substitutions in Eq. \eqref{eq:t1_t2_p_q_sub}, insert the result from Eq. \eqref{eq:pre_expo_t1t2_pq} and utilise the following expansion
\begin{equation}
\begin{aligned}
    -\frac{\RE(z)^2}{1+\tau} - \frac{\IM(z)^2}{1-\tau}& + \frac{z^2 + \bar{z}^2}{2 \tau} - N \left( \frac{t_1^2}{2 \tau} + \frac{i \bar{z} t_1}{\sqrt{N}\tau} - \ln(t_1) \right) -N \left( \frac{t_2^2}{2 \tau} - \frac{i z t_2}{\sqrt{N}\tau} - \ln(t_2) \right)  \\ 
    & \to  - N - \frac{y^2}{\kappa} + \frac{ip^3}{3} +ipx + \frac{iq^3}{3} +iqx + \frac{\frac{x^2}{2} + (p-q) y  + i (p+q) \kappa }{ N^{1/3} } + O(N^{-2/3}) \ ,
\end{aligned}
\end{equation}
which helps us to write down
\begin{equation}
\begin{aligned}
    Q_N \approx& \frac{1}{2 y} \sqrt{\frac{\kappa}{\pi}} \frac{N^{1/3}}{(2\pi i)^2} \int_{-\infty}^\infty dp \int_{-\infty}^\infty dq \, \frac{p-q}{i(p+q)} e^{ \frac{ip^3}{3} + ipx + \frac{iq^3}{3} + iqx} \left( 1+ \frac{\frac{x^2}{2} + (p-q) y  + i (p+q) \kappa }{ N^{1/3} } \right) \ .
\end{aligned}    
\end{equation}
We then realise that the only non-zero contribution comes from the term proportional to $(p-q)^2$. The integral can then be evaluated by changing $(p,q) \to i(p,q)$, deforming the contours of the $p$ and $q$ integrals to ${\cont}$ and recognising the derivatives of the Airy function, Eq. \eqref{eq:Ai_derivatives}. Returning to the mean self-overlap this yields
\begin{equation}
    \mathcal{O}_N^{\text{(eGinOE,c)}}(z) \approx \bigg[ \sqrt{\frac{\kappa}{\pi}} \exp { \frac{-y^2}{\kappa} } + \frac{\kappa}{2y}  \textup{erfc}\left( \frac{y}{\sqrt{\kappa}} \right)\bigg] \int_0^\infty dt \Big[ \textup{Ai}'(x+t)^2 - \textup{Ai}(x+t) \textup{Ai}''(x+t) \Big] \ ,
\end{equation}
and so we can now prove the result in Theorem \ref{thm:E_eGinOE_WNH_edge} by dividing the above expression by Eq. \eqref{eq:rho_eGinUE_WNH_edge}.

\subsubsection*{Acknowledgements}

We would like to thank Yan V. Fyodorov for interesting discussions and for offering feedback on this manuscript. We would also like to thank Wojciech Tarnowski for useful discussions. We would also like to thank Gernot Akemann for point out \cite{AB,AP} and their relevance to the question of universality. This research has been supported by the EPSRC Grant EP/V002473/1 “Random Hessians and Jacobians: theory and applications”. 

\appendix

\section{Proof of Proposition \ref{prop:rho_eGinOE_SNH_edge} - Density in the eGinOE} \label{app:rho_eGinOE_SNH}

In order to derive the density of complex eigenvalues at the edge in the eGinOE at strong non-Hermiticity, we start from the corresponding finite-$N$ equation, given in Eqs. \eqref{eq:rho_N_eGinOE} and \eqref{eq:P_N}. Initially, the prefactors can be simplified by noting that, in this regime, $\IM(z) \sim O(\sqrt{N})$. This can be used alongside the large argument behaviour of the complementary error function, $\erfc(x) \sim e^{-x^2}/(\sqrt{\pi} x)$ for $x\gg 1$, to show that, 
\begin{equation}
    \sqrt{\frac{2}{\pi}}\frac{\IM(z)}{1+\tau} \ \exp \left\{ \frac{\IM(z)^2 - \RE(z)^2}{1+\tau} \right\} \text{erfc}\left( \sqrt{\frac{2}{1-\tau^2}} \ | \IM(z) | \right) \approx \frac{1}{\pi} \sqrt{\frac{1 - \tau}{1 + \tau}} \exp{ - \frac{|z|^2 - \tau \Re(z^2)}{1 - \tau^2}} \ . 
\end{equation} 
To proceed from here, we introduce the integral representations of the Hermite polynomials given in Eq. \eqref{eq:He_k_int} and then utilise the series representation of the incomplete $\Gamma$-function in Eq. \eqref{eq:incmpl_Gamma}. After some manipulation, the density can be written as
\begin{equation}
\begin{aligned}
    \rho_N^{\textup{(eGinOE,c)}}(z) \approx & \frac{-1}{4 \pi^2 \tau } \frac{1}{\IM(z)} \sqrt{\frac{1 - \tau}{1 + \tau}} \exp{ - \frac{|z|^2 - \tau \Re(z^2)}{1 - \tau^2} + \frac{z^2 + \bar{z}^2}{2 \tau}} \\
    &\int_{-\infty}^\infty dt_1 \int_{-\infty}^\infty dt_2 \,(t_1 + t_2) \, \exp{ t_1 t_2 - \frac{t_1^2 + t_2^2}{2 \tau} + \frac{i}{\tau} \left( z t_2 - \bar{z} t_1 \right)  } \frac{\Gamma(N+1, t_1 t_2)}{\Gamma(N+1)}\ .
\end{aligned}
\end{equation}
We now make the same change of variables as defined in the vicinity of Eq. \eqref{eq:p_q_variables}, $\sqrt{N} p = t_1 + t_2$ and $\sqrt{N} q = t_1 - t_2$. Next, we utilise the integral representation of the incomplete $\Gamma$-function from Eq. \eqref{eq:incmpl_Gamma} and insert the parameterisation of the edge given in Eq. \eqref{eq:edge_xy_params}, yielding 
\begin{align}
    \rho&_N^{\textup{(eGinOE,c)}}(z) \approx \frac{-N^{N+3/2}}{8 \pi^2 \tau \Gamma(N)} \frac{1}{\IM(z)} \sqrt{\frac{1 - \tau}{1 + \tau}} e^{ - \frac{|z|^2 - \tau \Re(z^2)}{1 - \tau^2} + \frac{z^2 + \bar{z}^2}{2 \tau}} \int_1^\infty du \int_{-\infty}^\infty d p \int_{-\infty}^\infty dq \,  \frac{p(p^2 - q^2)}{4} e^{ - \frac{i \sqrt{N}w_x q + \sqrt{N} w_y p}{\tau} }  \nonumber\\
    & \exp{ N \left[ \frac{(1 - u)(p^2 - q^2)}{4} - \frac{p^2 + q^2}{4 \tau} - \frac{i(1+\tau) \cos(\theta)q}{\tau} - \frac{(1-\tau) \sin(\theta) p}{\tau} + \ln(\frac{u(p^2 - q^2)}{4}) \right]} \ . \label{eq:rho_edge_eGinOE_pre_exp}
\end{align}
As in the previous sections, the extremal point is located at $\left( p^*, q^*, u^* \right) = \left(-2 \sin(\theta), -2i \cos(\theta), 1 \right)$, hence we deform the contour and use the substitutions defined in Eq. \eqref{eq:pqu_subs}. Following the steps outlined in Appendix \ref{app:expo_SNH}, we find that
\begingroup
\allowdisplaybreaks
\begin{align}
    \rho_N^{\textup{(eGinOE,c)}}(z) \approx& \frac{N^{N}e^{-N}}{4 \pi^2 \tau \Gamma(N)} \frac{\sin(\theta)}{\IM(z)} \sqrt{\frac{1 - \tau}{1 + \tau}} e^{ - \frac{|z|^2 - \tau \Re(z^2)}{1 - \tau^2} + \frac{z^2 + \bar{z}^2}{2 \tau} + \frac{2\sqrt{N}(w_y \sin(\theta) - w_x \cos(\theta))}{\tau} - \frac{N(\tau + \cos(2\theta))}{\tau}} \nonumber \\
    & \int_0^\infty dv \, e^{-\frac{v^2}{2}} \int_{-\infty}^\infty d r \int_{-\infty}^\infty ds \exp{  - \frac{r^2( 1 - \tau \cos(2\theta))}{4 \tau} - \frac{w_y r}{\tau} + rv \sin(\theta)}  \label{eq:rho_eGinOE_SNH_coupled} \\ 
    &\exp{ - \frac{s^2( 1 - \tau \cos(2\theta))}{4 \tau} - \frac{i w_x s}{\tau} - i s v \cos(\theta) + i r s \cos(\theta) \sin(\theta) } \nonumber \ ,
\end{align}
\endgroup
where, in order to extract the leading order behaviour, we have used
\begin{equation}
    \frac{p(p^2 - q^2)}{4} \to \frac{1}{4} \left( \frac{r}{\sqrt{N}} - 2 \sin(\theta) \right) \left[ \left(\frac{r}{\sqrt{N}} - 2 \sin(\theta) \right)^2 - \left( \frac{s}{\sqrt{N}} - 2 i \cos(\theta)\right)^2 \right] = - 2 \sin(\theta) + O\left( N^{-1/2} \right) \ .
\end{equation}
The coupled Gaussian integrals over $r$ and $s$ can be evaluated using Eqs. \eqref{eq:CGI_start} to \eqref{eq:CGI_end}. Thus, after some additional manipulations on the first exponential, again involving Mathematica, this yields that
\begin{equation}
\begin{aligned}
    \rho_N^{\textup{(eGinOE,c)}}(z) \approx & \frac{N^N e^{-N}}{\pi \sqrt{1 + \tau^2 - 2 \tau \cos(2 \theta)} \Gamma(N)} \frac{\sin(\theta)}{\IM(z)} \sqrt{\frac{1 - \tau}{1 + \tau}} e^{- \frac{\eta^2 ( \tau (\tau^2 + 3) - (1 + 3 \tau^2) \cos(2 \theta) )}{\tau(1 - \tau^2)(1 + \tau^2 - 2 \tau \cos(2 \theta))} + \frac{\eta^2(\tau - \cos(2 \theta))}{\tau(1 + \tau^2 - 2 \tau \cos(2 \theta) )}} \\
    &\int_0^\infty dv \exp{- \frac{v^2}{2} + \frac{v^2 \tau(\tau - \cos(2 \theta))}{(1 + \tau^2 - 2\tau \cos(2\theta))} - \frac{ 2 v \eta }{\sqrt{1 + \tau^2 - 2\tau \cos(2\theta)}} } \ .
\end{aligned}
\end{equation}
If one now completes the square, exactly as in Eq. \eqref{eq:comp_square} and simplifies the prefactors by employing Stirling's formula and $\Im(z) \approx \sqrt{N} (1 - \tau) \sin(\theta)$, this produces 
\begin{equation}
\allowdisplaybreaks
\begin{aligned}
    \rho_N^{\textup{(eGinOE,c)}}(z) \approx & \frac{1}{\pi \sqrt{2\pi(1 - \tau^2)(1 + \tau^2 - 2 \tau \cos(2 \theta))}} \int_0^\infty dv \, e^{- \frac{1 - \tau^2}{2(1 + \tau^2 - 2 \tau \cos(2 \theta))} \left( v+ \frac{2 \eta \sqrt{1 + \tau^2 - 2 \tau \cos(2 \theta) }}{1 - \tau^2} \right)^2 } \\
    = & \frac{1}{\pi \sqrt{2 \pi} (1 - \tau^2)} \int_{\frac{2\eta}{ \sqrt{1- \tau^2}}}^\infty du \, \exp{-\frac{u^2}{2} } \ .
\end{aligned}
\end{equation}
Thus, one can complete the proof by using the definition of the complementary error function, $\erfc(x) = \sqrt{\frac{2}{\pi}}\int_{\sqrt{2}x}^\infty e^{-\frac{z^2}{2}} dz$, to see that, for large-$N$, the density is given by
\begin{equation}
    \rho_N^{\textup{(eGinOE,c)}}(z) \approx \frac{1}{2\pi (1 - \tau^2)} \erfc\left( \sqrt{\frac{2}{1 - \tau^2}} \eta \right) \ .
\end{equation}

\section{Expansion around Maximum of SNH Exponential}
\label{app:expo_SNH}

To derive our main results at strong non-Hermiticity, one must expand the function
\begin{align}
    H \equiv N \bigg[ \frac{(1 - u)(p^2 - q^2)}{4} - \frac{p^2 + q^2}{4 \tau} 
    - \frac{i(1+\tau) \cos(\theta)q}{\tau} - \frac{(1-\tau) \sin(\theta) p}{\tau} + \ln(\frac{u(p^2 - q^2)}{4}) \bigg] \ ,
\end{align}
which arises in the leading exponential of Eqs. \eqref{eq:O_edge_eGinUE_pre_exp} and \eqref{eq:O_edge_eGinOE_pre_exp},
in the vicinity of
\begin{equation}
    p = - 2 \sin(\theta) + \frac{r}{\sqrt{N}} \hspace{1cm} q = - 2 i \cos(\theta) + \frac{s}{\sqrt{N}} \hspace{1cm} u = 1 + \frac{v}{\sqrt{N}} \ , 
\end{equation}
as $N$ becomes large. Immediately, inserting the above substitutions and performing simple manipulations on the terms outside the logarithm, we find that
\begingroup
\allowdisplaybreaks
\begin{align}
    H \approx &  \frac{N}{\tau} \Big( (1 - \tau) \sin^2(\theta) - (1 + \tau) \cos^2(\theta)\Big) - \frac{(1-\tau)p^2}{4 \tau} - \frac{(1+\tau)q^2}{4 \tau}  - N \Bigg\{ \frac{1}{4}\left(1 + \frac{v}{\sqrt{N}}\right) \bigg\{ \left(\frac{r}{\sqrt{N}} - 2 \sin(\theta) \right)^2 \nonumber \\
    &  - \left( \frac{s}{\sqrt{N}} - 2 i \cos(\theta)\right)^2 \bigg]  - \ln( \frac{1}{4}\left(1 + \frac{v}{\sqrt{N}}\right) \bigg[ \left(\frac{r}{\sqrt{N}} - 2 \sin(\theta) \right)^2 - \left( \frac{s}{\sqrt{N}} - 2 i \cos(\theta)\right)^2 \bigg]) \Bigg\} \ .
    \label{eq:saddle_exp}
\end{align}
\endgroup
To evaluate the term in the curly brackets we write that
\begin{equation}
    \frac{1}{4}\left(1 + \frac{v}{\sqrt{N}}\right) \bigg[ \left(\frac{r}{\sqrt{N}} - 2 \sin(\theta) \right)^2 - \left( \frac{s}{\sqrt{N}} - 2 i \cos(\theta)\right)^2 \bigg] = 1 + \nu \ ,
\end{equation}
with 
\begin{equation}
    \nu = \frac{is \cos(\theta) - r \sin(\theta) + v}{\sqrt{N}} + \frac{{r}^2 - {s}^2 + 4v \big( is \cos(\theta) - r \sin(\theta)\big)}{4 N} + \frac{v({r}^2 - {s}^2)}{N^{3/2}} \ ,
\end{equation}
expanded in powers of $N$. Thus, if we Taylor expand the term of the form $1 + \nu - \ln(1 + \nu) \approx 1 +\frac{\nu^2}{2}$ in Eq. \eqref{eq:saddle_exp} and retain the leading order terms in $N$, we find that
\begin{equation}
    1 + \nu - \ln(1 + \nu) \approx 1 + \frac{1}{N} \left[ \frac{{r}^2 \sin^2(\theta) - {s}^2 \cos^2(\theta) + v^2}{2}  + i s v \cos(\theta) - rv \sin(\theta) - i r s \cos(\theta) \sin(\theta) \right] \ .
\end{equation}
Therefore, after some simple manipulations, involving trigonometric identities, one can see that 
\begin{equation}
\begin{aligned}
    H \approx - N - &\frac{N\big( \tau + \cos(2\theta)\big)}{\tau}   - \frac{v^2}{2} - \frac{{r}^2( 1 - \tau \cos(2\theta))}{4 \tau} \\
    &+ rv \sin(\theta) - \frac{{s}^2( 1 - \tau \cos(2\theta))}{4 \tau} - i sv \cos(\theta) + i r s \cos(\theta) \sin(\theta) \ .
    \label{eq:app_final_eq}
\end{aligned}
\end{equation}
This expansion is utilised in the proofs of the universality of $\rho(z)$ and $\mathcal{O}(z)$ at the edge in the regime of strong non-Hermiticity. As an example, it is used in the proof of Theorem \ref{thm:O_eGinOUE} in the eGinOE to arrive at Eq. \eqref{eq:O_eGinUE_SNH_coupled} from Eq.  \eqref{eq:O_edge_eGinUE_pre_exp}.

\end{document}